\documentclass[12pt]{article}
\newcommand{\filename}{2LMM-polymer-$\chi$}
\usepackage{authblk}
\usepackage{amsfonts}
\usepackage{amssymb}
\usepackage{amsmath}
\usepackage{amsthm}
\usepackage{framed}
\usepackage{graphicx}
\usepackage{siunitx}
\usepackage{subcaption}

\usepackage[nocompress]{cite}

\usepackage{enumitem}
\usepackage{geometry}
\usepackage{url}
\usepackage{booktabs}

\usepackage{multirow}

\usepackage{mdframed}
\usepackage{mathtools}
\usepackage{algorithm}  
\usepackage{algorithmic}

\newcommand{\bbC}{\mathbb{C}}
\newcommand{\bbR}{\mathbb{R}}
\newcommand{\bbZ}{\mathbb{Z}}
\newcommand{\anC}{\langle \mathbb{C} \rangle}

\newcommand{\GC}{G_{\mathrm{C}}}
\newcommand{\VC}{V_{\mathrm{C}}}
\newcommand{\EC}{E_{\mathrm{C}}}

\newcommand{\Ez}{E_{(0/1)}}
\newcommand{\Ew}{E_{(\geq 1)}}
\newcommand{\Et}{E_{(\geq 2)}}
\newcommand{\Eew}{E_{(=1)}}

\newcommand{\ta}{{\tt a}}
\newcommand{\tb}{{\tt b}}

\newcommand{\tC}{{\tt C}}
\newcommand{\tO}{{\tt O}}
\newcommand{\tN}{{\tt N}}
\newcommand{\tS}{{\tt S}}
\newcommand{\tH}{{\tt H}}
\newcommand{\tP}{{\tt P}}

\newcommand{\VH}{V_{\tH}}

\newcommand{\calG}{\mathcal{G}}

\newcommand{\ylb}{\underline{y}^*}
\newcommand{\yub}{\overline{y}^*}

\newcommand{\val}{\mathrm{val}}
\newcommand{\h}{\mathrm{ht}}
\newcommand{\dg}{\mathrm{dg}}

\newcommand{\ex}{\mathrm{ex}}
\newcommand{\inte}{\mathrm{int}}
\newcommand{\lnk}{\mathrm{lnk}}

\newcommand{\Vleaf}{V_{\mathrm{leaf}}}
\newcommand{\Eleaf}{E_{\mathrm{leaf}}}
\newcommand{\Elnk}{E^\lnk}

\newcommand{\chiaoki}{{\sc Chi-Aoki}}
\newcommand{\chinistane}{{\sc Chi-Nistane}}
\newcommand{\chijsol}{{\sc Chi-JOCTA}}
\newcommand{\jocta}{{J-OCTA}}

\newcommand{\Rt}{\mathrm{R}^2}

\newcommand{\molinfer}{{\tt mol-infer}}

\newcommand{\cnt}{\mathrm{cnt}}  

\newcommand{\nlnk}{\mathrm{n}^\mathrm{lnk}}

\newcommand{\lf}{\mathrm{lf}}

\newcommand{\ttH}{{\tt H}}  
\newcommand{\ttC}{{\tt C}}  
\newcommand{\ttO}{{\tt O}}  
\newcommand{\ttN}{{\tt N}}  
\newcommand{\ttP}{{\tt P}}  
  
\newcommand{\ttCl}{{\tt Cl}}  
\newcommand{\ttS}{{\tt S}}

\newcommand{\oH}{\overline{{\tt H}}}  

\newcommand{\Z}{\mathbb{Z}}  
  
\newcommand{\C}{\mathbb{C}}  
\newcommand{\Co}{\mathbb{C}}  
\newcommand{\fr}{\mathrm{fr}}    

\newcommand{\anpsi}{\langle \psi \rangle}

\newcommand{\dcp}{\mathrm{dcp}}

\newcommand{\sint}{\sigma_\mathrm{int}} 
\newcommand{\sce}{\sigma_\mathrm{ce}}

\newcommand{\Gac}{\Gamma_\mathrm{ac}}

\newcommand{\typ}{\mathrm{t}}

\newcommand{\Ldg}{\Lambda_{\mathrm{dg}}}

\newcommand{\fc}{\mathrm{fc}}

\newcommand{\nint}{\mathrm{n}^\mathrm{int}}

\newcommand{\cs}{\mathrm{cs}}
\newcommand{\ch}{\mathrm{ch}}

\newcommand{\na}{\mathrm{na}}

\newcommand{\acC}{\mathrm{ac}_\mathrm{C}}

\newcommand{\ns}{\mathrm{ns}}

\newcommand{\ec}{\mathrm{ec}}
\newcommand{\ac}{\mathrm{ac}}

\newcommand{\bl}{\mathrm{bl}}

\newcommand{\bd}{\mathrm{bd}}

\newcommand{\UB}{\mathrm{UB}}
\newcommand{\LB}{\mathrm{LB}}

\begin{document} 

\begin{center}
   {\Large\bf 
Towards Environment-Sensitive Molecular Inference via Mixed Integer Linear Programming
  }
\end{center} 
\begin{center}
Jianshen Zhu$^1$, 
Mao Takekida$^1$,
Naveed Ahmed Azam$^2$, 
Kazuya Haraguchi$^{1}$, 
Liang Zhao$^3$, 
 and  
 Tatsuya Akutsu$^4$ 
\end{center} 
%
%
{\small 
$^1$Graduate School of Informatics, Kyoto University, Kyoto 606-8501, Japan\\
$^2$Department of Mathematics, Quaid-i-Azam University, Islamabad 45320, Pakistan\\
$^3$Graduate School of Advanced Integrated Studies in Human Survivability   (Shishu-Kan),  
  Kyoto University, Kyoto 606-8306, Japan \\
$^4$Bioinformatics Center,  Institute for Chemical Research, 
  Kyoto University, Uji 611-0011, Japan 
}

\begin{quote}  
{\bf Abstract}\\  
Traditional QSAR/QSPR and inverse QSAR/QSPR methods often assume that chemical properties
are dictated by single molecules, overlooking the influence of molecular interactions and environmental factors.
In this paper, we introduce a novel QSAR/QSPR framework that can capture the combined effects
of multiple molecules (e.g., small molecules or polymers)
and experimental conditions on property values.
We design a feature function to integrate the information of multiple molecules and the environment.
Specifically, for the property Flory-Huggins $\chi$-parameter, 
which characterizes the thermodynamic properties between
the solute and the solvent, 
and varies in temperatures, we demonstrate through computational experimental results that
our approach can achieve a competitively high learning performance
compared to existing works on predicting $\chi$-parameter values, while inferring the solute polymers with up to 50 non-hydrogen 
atoms in their monomer forms in a relatively short time.
A comparison study with the simulation software \jocta\ demonstrates that the polymers inferred by our methods are of high quality.



\end{quote}

\section{Introduction}\label{sec:introduction}

\noindent

In recent years, there have been extensive studies focusing on the design of novel molecules using 
machine learning techniques~\cite{Lo:2018aa, Tetko:2020aa}.
Molecular design with the aid of computational methods
has a long history in the field of cheminformatics,
and are commonly studied under the name of \emph{quantitative structure activity/property relationship} 
(QSAR/QSPR)~\cite{Cherkasov:2014aa, Skvortsova:1993aa}, and \emph{inverse quantitative structure
activity/property relationship} (inverse QSAR/QSPR)~\cite{Ikebata:2017aa, Miyao:2016aa, Rupakheti:2015aa}.
QSAR/QSPR models aim to predict chemical activities based on molecular
structures~\cite{Cherkasov:2014aa}, whereas
inverse QSAR/QSPR models focus on inferring molecular structures that exhibit specific
chemical activities/properties~\cite{Ikebata:2017aa, Miyao:2016aa, Rupakheti:2015aa}.
Traditionally, molecular structures are represented as undirected graphs, referred to as
\emph{chemical graphs}, and typically encoded as vectors of real numbers called \emph{descriptors} or
\emph{feature vectors} in most of the existing QSAR/QSPR and inverse QSAR/QSPR studies.
A typical approach for inverse QSAR/QSPR involves inferring feature vectors from
 given chemical activities and subsequently reconstructing chemical graphs from 
 these vectors~\cite{Ikebata:2017aa, Miyao:2016aa, Rupakheti:2015aa}. 
 Recently,
 artificial neural network (ANN) and other deep learning techniques 
 have been increasingly utilized in inverse QSAR/QSPR studies,
due to the availability of generative models~\cite{Ghasemi:2018aa, Kipf:2016aa, Gomez-Bombarelli:2018aa, De-Cao:2018aa, Madhawa:2019aa, Du:2022aa}.

In the last few years, significant breakthroughs have been made in the field of computational molecular modeling,
highlighted with the release of AlphaFold~3~\cite{Jumper:2021aa, Abramson:2024aa},
which demonstrates remarkable accuracy in predicting protein structures using state-of-the-art deep learning models
using 3D spatial information.
However, the inference of small molecules remains an important task,
as they play an important role in drug design and related applications.
For instance, the ChEMBL database~\cite{Zdrazil:2023aa} contains more than
$2 \times 10^6$ small drug-like molecules. While 3D spatial information can reveal critical molecular insights,
it is often unavailable in such databases.
Consequently, the use of only 2D structural information, such as chemical graphs and topological features,
remains an essential approach in QSAR/QSPR and inverse QSAR/QSPR studies.


Most of the existing QSAR/QSPR and inverse QSAR/QSPR models 
are developed under the assumption
that a chemical property value is determined by just one
molecule, 
and the environment such as temperature and pressure when the value is measured,
 is neglected more or less.
 Meanwhile, some important properties depend on interactions between multiple molecules
 and are sensitive to factors like
temperature, pressure, and frequency, collectively referred to as the
{\em environment} in this study.
For example, permittivity depends on the frequency, magnitude, and direction of the applied field~\cite{Landau:2013aa}, 
and dissipation factor varies on the frequency of electrical signals~\cite{Valentine:2019aa}.
One important value that characterizes the thermodynamic properties between the solute and the solvent, Flory-Huggins $\chi$-parameter,
is also related to two molecules, and is found depending on temperature, pressure, and some other factors~\cite{Aoki:2023aa}.

To fill this research gap,
in this paper, we extend the 
existing framework \molinfer~\cite{Chiewvanichakorn:2020aa, Shi:2021aa, Zhu:2022ad, Ido:2024aa} 
to include such important cases by integrating the information of multiple molecules and the 
experimental environmental factors into one feature vector.
In~\cite{Ido:2024aa}, the authors 
considered the frequencies when the values are observed as descriptors when constructing a prediction function for the property
permittivity, but did not discuss this issue formally.
As an example of the extended framework, we particularly
consider the task of inferring molecules with desired $\chi$-parameter values under specific environments.
Various models have been developed to 
predict Flory-Huggins $\chi$-parameter values 
computationally~\cite{Hildebrand:1950aa, Lindvig:2002aa, Orwoll:2007aa},
but these methods are either known for their low prediction performance or high computational cost~\cite{Aoki:2023aa, Nistane:2022aa, Orwoll:2007aa, Wolf:2011aa}.
Also, as far as we know, 
there is no work on designing molecules with specified $\chi$-parameter values.
We design a feature vector for the solute-solvent pair $(\bbC_1, \bbC_2)$ and manage to include the temperature information into it 
in several different ways, and then apply three different machine learning methods to two data sets on empirical $\chi$-parameter values 
and one data set on simulation-based $\chi$-parameter values. 
The experimental results highlight the model's ability
to achieve a competitively high learning performance, comparable to or exceeding existing studies~\cite{Aoki:2023aa, Nistane:2022aa}
on predicting $\chi$-parameter values.
Also, focusing on the situation of inferring a solute $\bbC_1$ when given a fixed solvent $\bbC_2$ and a specified temperature $T$,
we successfully infer
polymers with up to 50 non-hydrogen atoms in their monomer forms in a reasonable time.
A comparative study to the simulation software \jocta~\cite{JOCTA} demonstrates that the polymers
generated by our proposed methods are generally of good quality.

We organize the paper as follows.
Section~\ref{sec:preliminary} reviews some basic concepts and terminologies on graphs,
the framework \molinfer\ to infer a molecule with some desired property values,
and a modeling of chemical compounds.
Section~\ref{sec:formulation} describes how we extend the framework for multiple molecules and experimental environmental factors.
Section~\ref{sec:experiment} presents the results of 
computational experiments conducted on data sets of $\chi$-parameter.
Section~\ref{sec:conclude} concludes the paper.
Additional details are included in the Appendix.
All the program codes and experimental results are accessible at {\url{https://github.com/ku-dml/mol-infer/tree/master/chi-parameter}}.

 \section{Preliminary}\label{sec:preliminary}
We give some notions and terminologies on graphs in Section~\ref{sec:graphs}
and review the framework \molinfer\ in Section~\ref{sec:frame_all},
a modeling of chemical compounds in Section~\ref{sec:chemical_model},
and the two-layered model, the standard model in \molinfer, in Section~\ref{sec:2LM}.
Some necessary modifications introduced by Ido~et~al.~\cite{Ido:2024aa} when the molecule is a polymer 
will be covered in Section~\ref{sec:polymer}.


Let $\bbR$, $\bbR_+$, $\bbZ$  and $\bbZ_+$ 
represent the sets of reals,  non-negative reals, 
integers, and non-negative integers, respectively.
For two integers $a$ and $b$ such that $a \leq b$, $[a,b]$ is defined as the set of 
integers $i$ such that $a\leq i\leq b$.

\subsection{Graphs}\label{sec:graphs}

When referring to a {\em graph} $G$, it is assumed that $G$ is a connected and simple graph.
The sets of vertices and edges of a given graph $G$ are denoted
by $V(G)$ and $E(G)$, respectively.
For any vertex $v\in V(G)$, we denote the set of its neighbors in $G$ by $N_G(v)$,
and the {\em degree} $\mathrm{deg}_G(v)$ of $v$ is
$\mathrm{deg}_G(v)=|N_G(v)|$.

A vertex designated in a graph $G$ is called a {\em root},
and a graph with such a vertex is referred to as a {\em rooted graph}. 
 For a graph $G$ (possibly rooted),
 a {\em leaf-vertex} is a non-root vertex  $v$ with degree 1.
 For any subset $V'\subseteq V(G)$, 
the graph $G-V'$ is obtained by removing all vertices in $V'$ along with any edges incident to them. 
An edge $uv$ incident to a leaf-vertex $v$ is called a {\em leaf-edge}.
 We denote the sets of leaf-vertices and leaf-edges in $G$ by $\Vleaf(G)$ and $\Eleaf(G)$, respectively.
 For a graph $G$ (possibly rooted),
 a sequence of graphs $G_i, i\in \mathbb{Z}_+$ is defined by iteratively removing all leaf-vertices
 $i$ times as follows:
\[ G_0:=G; ~~ G_{i+1}:=G_i - \Vleaf(G_i). \]
A vertex $v$ is called a {\em tree vertex} if $v\in \Vleaf(G_i)$
for some $i\geq 0$. 
We define the {\em height} $\h(v)$ of a tree vertex $v\in \Vleaf(G_i)$ to be $i$; 
and for a non-tree vertex $v$ adjacent to a tree vertex, we define the height
$\h(v)$ to be $\h(u)+1$, 
where $u$ is the tree vertex with the maximum height $\h(u)$ among those adjacent to $v$.
The heights of other vertices are left undefined. 
Finally, the {\em height} $\h(T)$ of a rooted tree $T$ is defined
to be the maximum of $\h(v)$ among all vertices $v\in V(T)$.

\subsection{\molinfer : An Inverse QSAR/QSPR Framework Based on Machine Learning and MILP}\label{sec:frame_all}

\begin{figure}[t!]
\begin{center}
\includegraphics[width=.8\columnwidth]{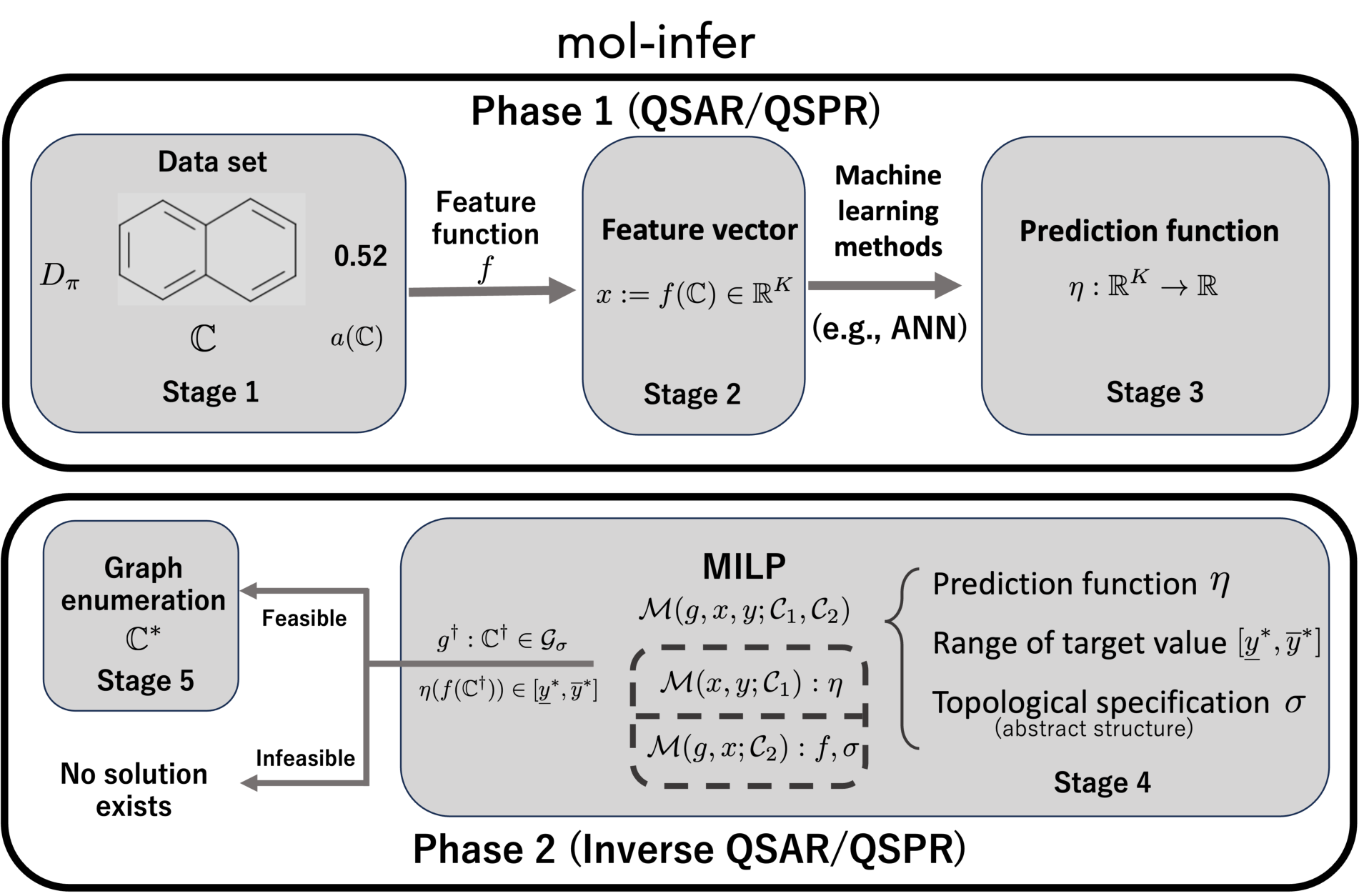}
\end{center}
\caption{An illustration of the two-phase framework \molinfer.} 
\label{fig:framework}
\end{figure}

The computation process of an artificial neural network (ANN)
with ReLU activation functions can be 
represented through a mixed integer linear programming (MILP) formulation,
as demonstrated by Akutsu and Nagamochi~\cite{Akutsu:2019aa}.
%
Based on this concept, 
a two-phase inverse QSAR/QSPR framework,
 called \molinfer,
has been proposed and subsequently refined~\cite{Chiewvanichakorn:2020aa,Zhang:2022aa, Shi:2021aa, Zhu:2022ad, Ido:2024aa, Zhu:2023aa},
as depicted in Figure~\ref{fig:framework}.
This framework mainly establishes on using the \emph{mixed integer linear programming} (MILP) formulation
to simulate the computational process of machine learning methods and describe 
the necessary and sufficient conditions to ensure such a chemical graph exists,
utilizing only 2D structural information.
The advantage of \molinfer\ compared to other methods is that it
guarantees both optimality and exactness.
Here, optimality refers to the quality of the solution in addressing 
the inverse problem of learning methods,
while exactness ensures that the solution corresponds to a valid chemical graph.
This framework was first introduced for general molecules~\cite{Chiewvanichakorn:2020aa,
Shi:2021aa, Zhu:2022ad} and then extended to
polymers recently~\cite{Ido:2024aa}.
This subsection provides an overview of the core ideas for \molinfer\ for completeness.

\subsubsection{Phase~1}
Phase~1 is the QSAR/QSPR phase, aiming to construct a prediction function between chemical compounds and their observed property values, 
and consisting of three stages. 
Here we denote $\calG$ the set of all possible chemical graphs.
\begin{itemize}
\item[-] Stage~1: Given a chemical property $\pi$, we collect a data set $D_\pi \subseteq \calG$ of chemical graphs such that
for every chemical graph $\bbC \in D_\pi$, the observed value $a(\bbC)$ of property $\pi$ 
is available.
\item[-] Stage~2: A feature function $f: \calG \to \bbR^K$ ($K$ is a positive integer) is defined. 
This feature function consists of only graph-theoretic descriptors, 
mainly based on the local graph-theoretic structures of the chemical graph
 so that $f$ is tractable by MILP formulations in Phase~2. 
\item[-] Stage~3: A prediction function $\eta$ is constructed by some machine learning methods
in order to produce an output $y=\eta(x)\in\bbR$ based
on the feature vector $x=f(\bbC)\in\bbR^K$ for each $\bbC \in D_\pi$.
\end{itemize}

\subsubsection{Phase~2}
Phase~2 is devoted to the inverse QSAR/QSPR phase, 
designed to infer chemical graphs with a specified property value
based on the prediction function $\eta$ constructed in Phase~1. 
It consists of two stages.

\begin{itemize}
\item[-] Stage~4: Given a set of rules called topological specification $\sigma$ (see Section~\ref{sec:2LM} for more details) that specifies
the desired structure of the inferred chemical graphs, and a desired range $[\ylb, \yub]$ of the target value, 
Stage 4 is to infer a chemical graphs $\bbC^\dagger$ that satisfies the rules $\sigma$ and
$\eta(f(\bbC^\dagger)) \in [\ylb, \yub]$.
To achieve this,
an MILP formulation $\mathcal{M}(g,x,y;\mathcal{C}_1,\mathcal{C}_2)$ is formulated, 
which consists of two parts:
\begin{itemize}
\item[(i)] $\mathcal{M}(x,y;\mathcal{C}_1)$: the computation process of $y := \eta(x)$ from a vector $x \in \bbR^K$; and
\item[(ii)] $\mathcal{M}(g,x;\mathcal{C}_2)$: that of $x := f(\bbC)$ and the constraints for $\bbC \in \calG_\sigma$,
\end{itemize}
where $\calG_\sigma$ denotes the set of all chemical graphs satisfying $\sigma$.
We solve the MILP $\mathcal{M}(g,x,y;\mathcal{C}_1,\mathcal{C}_2)$ 
for a given $\sigma$ and $[\ylb, \yub]$
to find
a feature vector $x^* \in \bbR^K$ and a chemical graph $\bbC^\dagger$
such that $f(\bbC^\dagger) = x^*$ and $\eta(x^*) \in [\ylb, \yub]$.
In the case that the MILP is infeasible, it indicates that no chemical graph in $\calG_\sigma$ satisfies the specified demand.
\item[-] Stage~5:
The final stage is to generate the isomers of the inferred chemical graphs $\bbC^\dagger$
by using a dynamic programming-based graph enumeration algorithm developed by Zhu~et~al.~\cite{Zhu:2022ad}. 
A {\em chemical isomer} of $\bbC^\dagger$ under
a topological specification $\sigma$ is defined as 
a chemical graph $\bbC^*$  such that
$f(\bbC^*)=f(\bbC^\dagger)$ and $\bbC^*\in \calG_\sigma$.
This graph enumeration algorithm operates by decomposing $\bbC^\dagger$
into trees and generating their isomers respectively. These isomers
are then combined to produce a set of chemical isomers $\bbC^*$ that belong to the desired
chemical graph space $\calG_\sigma$ and have exactly the same feature vector as $\bbC^\dagger$.
\end{itemize}

\subsection{Modeling of Chemical Compounds}\label{sec:chemical_model}

This subsection reviews a modeling of chemical compounds 
introduced by Zhu~et~al.\cite{Zhu:2022ad}.
Let $\Lambda$ represent the set of chemical elements;
for example,  $\Lambda=\{\tH,  \tC, \tO, \tN, \tP, \tS_{(2)}, \tS_{(4)}, \tS_{(6)}\}$. 
Here, elements $\ta$ with multiple valence states are distinguished with a suffix, i.e., 
we denote an element $\ta$ with valence $i$ as $\ta_{(i)}$.


A chemical compound $\bbC$ is represented as a {\em chemical graph}, which is defined as
a triplet $\bbC=(H,\alpha,\beta)$, where
$H$ is a graph $H$,
$\alpha:V(H)\to \Lambda$ assigns chemical elements to vertices, and
$\beta:E(H)\to [1,3]$ assigns bond multiplicities to edges.
Two chemical graphs $(H_1, \alpha_1, \beta_1)$ and $(H_2,\alpha_2,\beta_2)$ are
{\em isomorphic} if there exists
an isomorphism $\phi$,
i.e., a bijection $\phi: V(H_1)\to V(H_2)$
such that
 $uv\in E(H_1), \alpha_1(u)=\ta, \alpha_1(v)=\tb, \beta_1(uv)=m$
if and only if
 $\phi(u)\phi(v) \in E(H_2), \alpha_2(\phi(u))=\ta, 
 \alpha_2(\phi(v))=\tb, \beta_2(\phi(u)\phi(v))=m$. 
 If $H_1$ and $H_2$ are rooted graphs with roots $r_1$ and $r_2$, respectively,
 the chemical graphs are considered
{\em rooted-isomorphic} 
if there exists an isomorphism $\phi$ such that $\phi(r_1)=r_2$ also holds. 
  
For a chemical graph  $\bbC=(H,\alpha,\beta)$, 
  let  $V_{\ta}(\bbC)$ ($\ta\in \Lambda$)
represent the set of vertices $v\in V(H)$ such that $\alpha(v)=\ta$. 
The {\em hydrogen-suppressed chemical graph} of $\bbC$, denoted as $\anC$,
is obtained by removing all vertices $v\in \VH(\bbC)$ from $H$.

\subsubsection{Two-layered Model}\label{sec:2LM}
Shi~et~al.~\cite{Shi:2021aa} introduced the
two-layered model for chemical graphs
to efficiently capture the graph-theoretic information and develop descriptors based on it.
Here, we summarize the key concepts of this model for completeness.
 
\begin{figure}[t!] \begin{center}
\includegraphics[width=.90\columnwidth]{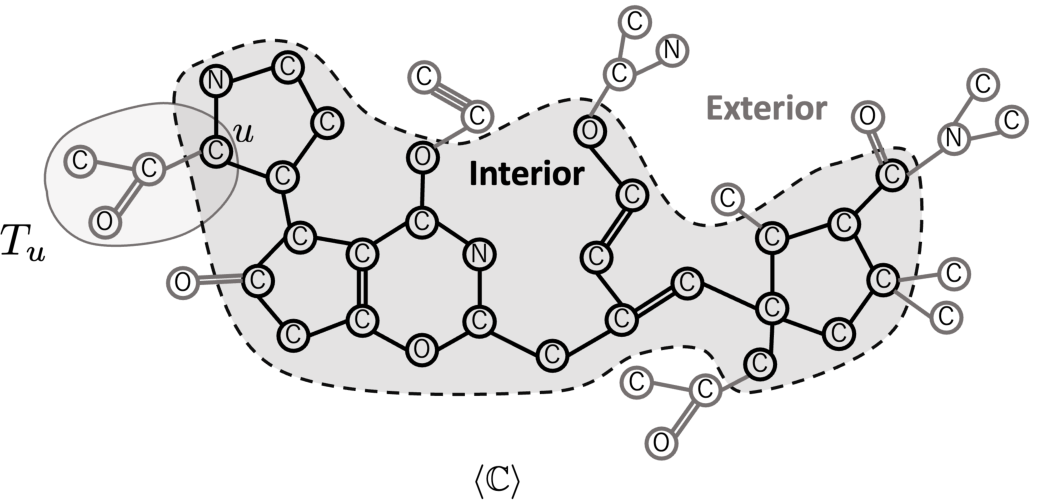}
\end{center}
\caption{
An illustration of the two-layered model.
The interior region is represented by the shaded area enclosed by black dashed lines,
while the remaining parts form the exterior.
$T_u$ is the chemical tree rooted at $u$ and is outlined by a thin gray line.
 }
\label{fig:two_layer} \end{figure}

Consider a chemical graph $\bbC=(H,\alpha,\beta)$ and an integer $\rho \geq 1$,
referred to as the {\em branch-parameter}.
For this study, the standard value of $\rho = 2$ is used.
The {\em two-layered model} of $\bbC$ is a partition of
 the hydrogen-suppressed chemical graph $\anC$ into
two regions: the ``interior'' and the ``exterior'' based on the branch-parameter $\rho$.
A vertex $v \in V(\anC)$ (resp., an edge $e \in E(\anC)$) of $\bbC$
is classified as an {\em exterior-vertex} (resp., {\em exterior-edge})
if $\h(v)< {\rho}$ (resp., $e$ is incident to an exterior-vertex).
We denote 
the sets of exterior-vertices and exterior-edges of $\bbC$
by $V^\ex(\bbC)$ and $E^\ex(\bbC)$, respectively. 
The remaining vertices and edges, defined as
$V^\inte(\bbC)=V(\anC)\setminus  V^\ex(\bbC)$ and 
$E^\inte(\bbC)=E(\anC)\setminus E^\ex(\bbC)$,
are called  {\em interior-vertices} and {\em interior-edges}, respectively.
Notice that the set  $E^\ex(\bbC)$  forms 
a collection of connected graphs, each can be treated as a rooted tree $T$ with the root being 
the vertex $v\in V(T)$ with the maximum height $\h(v)$. 
Let $\mathcal{T}^\ex(\anC)$ denote 
the set of these rooted trees in $\anC$. 
The {\em interior} of $\bbC$ is defined to be the subgraph
 $(V^\inte(\bbC),E^\inte(\bbC))$ of $\anC$. 
See Figure~\ref{fig:two_layer} for an example.
 


For each interior-vertex $u\in V^\inte(\bbC)$,
let $T_u\in \mathcal{T}^\ex(\anC)$ represent the chemical tree rooted at $u$
(where $T_u$ may consist solely of the vertex $u$).
The {\em $\rho$-fringe-tree} $\bbC[u]$  is defined
as the chemical rooted tree obtained by restoring the hydrogens which are originally attached
to $T_u$ in $\bbC$.

For a given integer $K$, a feature vector $f(\bbC)$ for a chemical graph $\bbC$
is defined by a {\em feature function} $f$ which comprises $K$ descriptors 
based on the two-layered model.
A comprehensive list and detailed explanation of the feature function $f$ used in this study
 can be found in Appendix~\ref{sec:descriptor}.

Furthermore, in order to allow the usage of domain knowledge for inference of chemical graphs,
such as some abstract structures or limits on the available 2-fringe-trees,
a set of rules called {\em topological specification} is used.
A topological specification includes the following components:
\begin{itemize}
\item[-] A {\em seed graph} $\GC$, which serves as an abstract form of the target chemical graph $\bbC$.
\item[-] A set $\mathcal{F}$ of chemical rooted trees, which serve as candidates 
for the tree  $\bbC[u]$ rooted at each interior-vertex $u$ in $\bbC$.
\item[-] Lower and upper bounds that constrain the number of various components 
in the target chemical graph, such as the interior-vertices, 
double/triple bonds, and chemical elements in $\bbC$. 
\end{itemize}

\begin{figure}[t!]
\begin{center} 
 \includegraphics[width=.92\columnwidth]{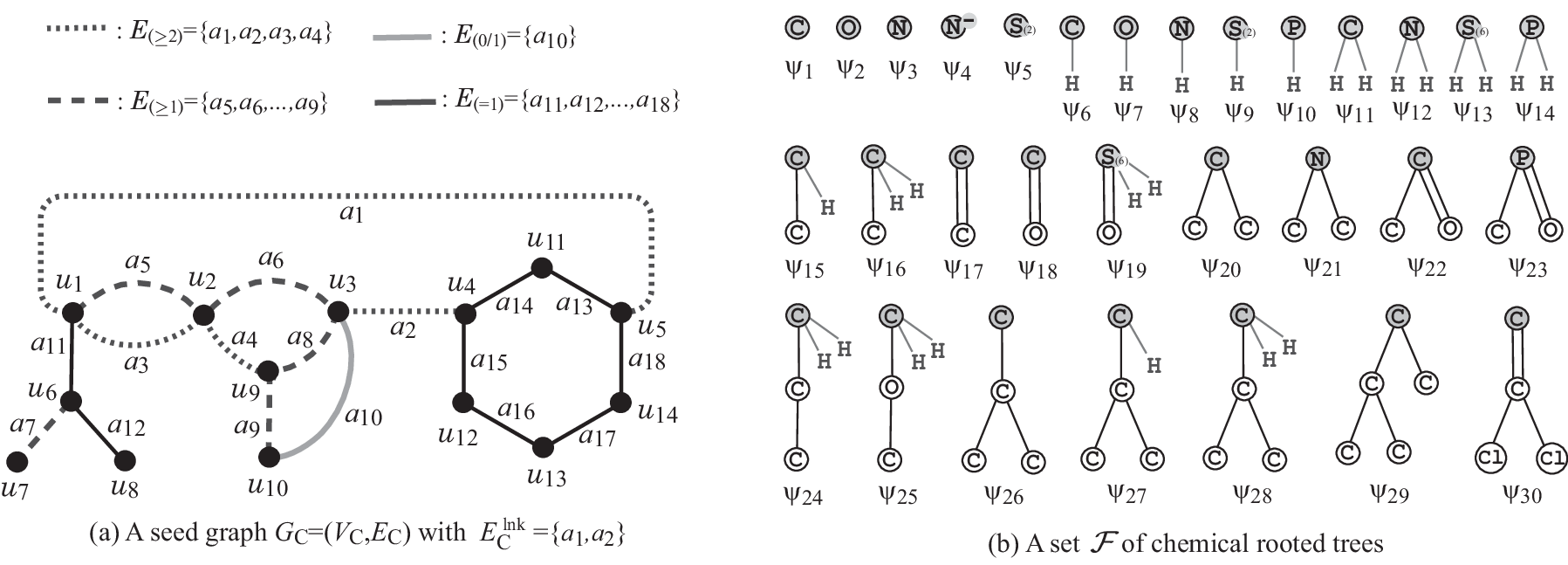}
\end{center}
\caption{(i) A seed graph $G_{\mathrm{C}}$ for $I_a$; (ii) A set $\mathcal{F}$ of chemical rooted trees.
The figure is adapted from~\cite{Ido:2024aa}.
}
\label{fig:seed_graph_a}  
\end{figure} 

Figure~\ref{fig:seed_graph_a} illustrates one example of topological specification.
We refer~\cite{Zhu:2022ad} and Appendix~\ref{sec:specification} for a more detailed description of the topological specification.

\subsubsection{Modeling of Polymers}\label{sec:polymer}

In this subsection, we review the way of representing a polymer as a form of monomer that is proposed by Ido~et~al.~\cite{Ido:2024aa},
and the necessary modification for polymers in the two-layered model.

\begin{figure}[t!] \begin{center}
\includegraphics[width=.96\columnwidth]{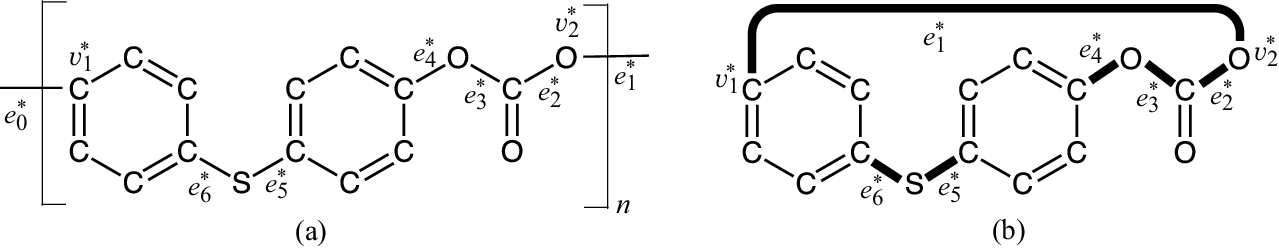}
\end{center}
\caption{(a) The repeating unit of the polymer thioBis(4-phenyl)carbonate,
where $v^*_1$ and $v^*_2$ are the connecting-vertices
and $e^*_0$ and $e^*_1$ are the connecting-edges;
(b) The monomer representation of the polymer in (a), where 
$v^*_1$ and $v^*_2$ are the connecting-vertices
and the link-edges are depicted with thick lines.
The figure is adapted from~\cite{Ido:2024aa}.}
\label{fig:polymer_example}  \end{figure}

For polymers,
we mainly focus on the case of homopolymer, i.e., a linear sequence
of identical repeating units connected by two specific edges, $e^*_0$ and $e^*_1$,
such that two adjacent units in the sequence are joined with them.
The two edges are referred to as the {\em connecting-edges},
and the two vertices incident 
to the two connecting-edges are called the {\em connecting-vertices}.  
An example of these concepts can be found in Figure~\ref{fig:polymer_example}(a).
 
We call an edge $e$ a  {\em link-edge} 
 in a repeating unit of a polymer if it is traversed by every path 
connecting $e^*_0$ and $e^*_1$,
and denote the set of link-edges in $\bbC$ by $\Elnk(\bbC)$.
For instance, in  Figure~\ref{fig:polymer_example}(a), the link-edges 
 are $e^*_2,e^*_3,\ldots,e^*_6$. 
Following~\cite{Ido:2024aa},
we treat the two connecting-edges as a single edge $e^*_1$ 
to simplify the representation of the polymer,
as illustrated in Figure~\ref{fig:polymer_example}(b). 
The resulting graph is called
the {\em monomer representation} of the polymer,
and edge $e^*_1$ is also called a {\em link-edge}.
In what follows, we represent polymers by their monomer representations $\bbC$.


The link-edges and connecting-vertices are both used in the feature function defined for polymers.
See Appendix~\ref{sec:descriptor} for more details about this.
Also, we specify the set of link-edges in the seed graph $\GC$ and the lower and upper bounds on the number
of components such as the  link-edges and connecting-vertices
in the topological specification used for a polymer.
See Appendix~\ref{sec:specification} for a more detailed explanation of the topological specification for polymers.


\section{An Extended Framework for Multiple Molecules and Environments}\label{sec:formulation}

In this section, we describe how we extend the framework \molinfer\ 
to take multiple molecules into consideration
and include information about the environment.
The extended framework still consists of two phases like \molinfer, Phase~1 for the QSAR/QSPR phase,
and Phase~2 for the inverse QSAR/QSPR phase.

\subsection{Phase~1}\label{sec:formulation_phase1}

\begin{figure}[t!]
\begin{center} 
 \includegraphics[width=.95\columnwidth]{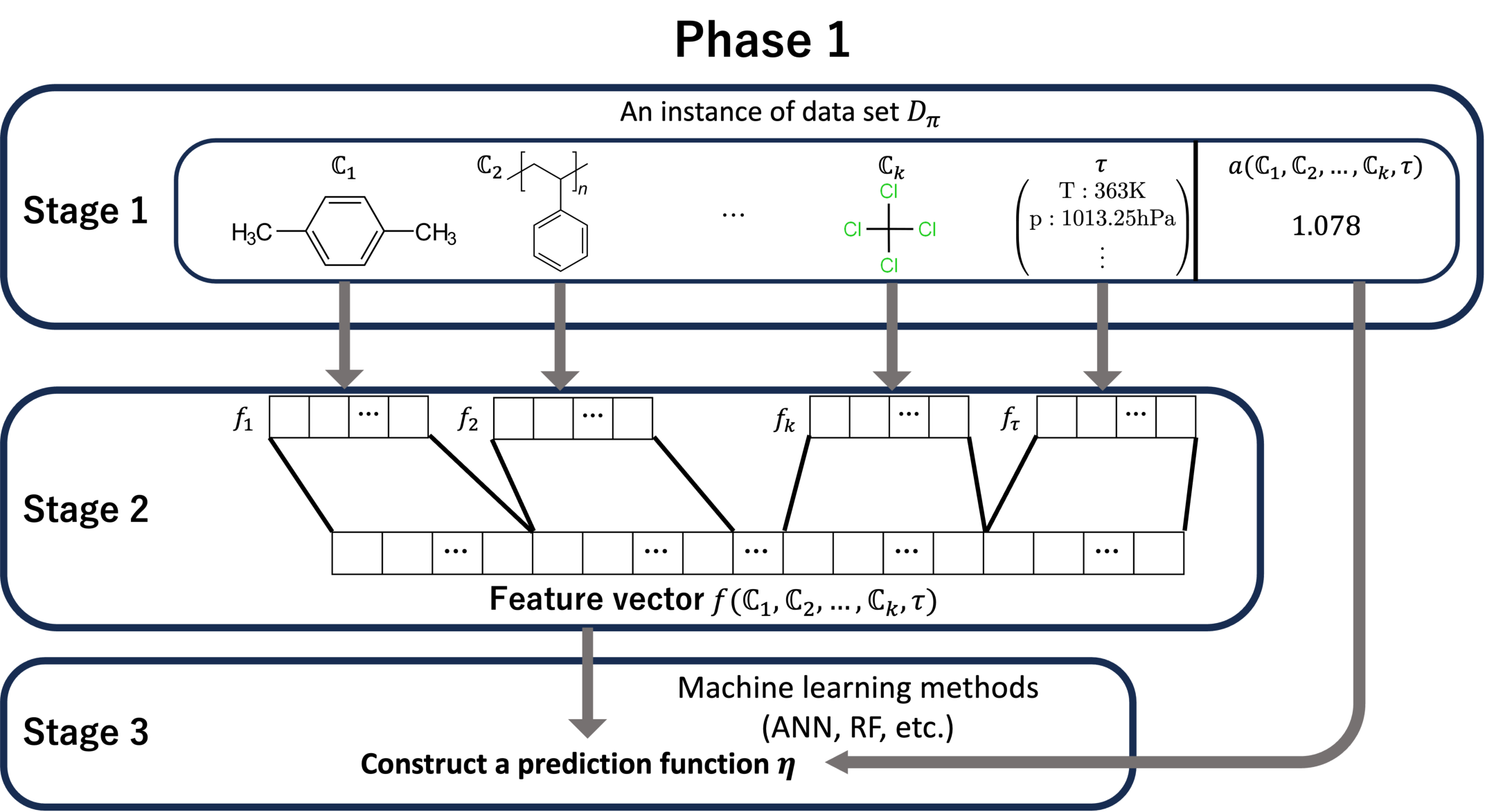}
\end{center}
\caption{An illustration of Phase~1 of the extended framework.
}
\label{fig:figure_1}  
\end{figure} 

Figure~\ref{fig:figure_1} shows Phase~1 of the extended framework.

Let $k$ and $m$ be two positive integers.
In Stage~1,
given a property $\pi$,
we define an instance of $\pi$ to be a $(k+1)$-tuple $(\bbC_1, \bbC_2, ..., \bbC_k, \tau)$,
where
$k$ denotes the number of chemical graphs $\bbC_1, \bbC_2, ..., \bbC_k$ involved in determining the value of the property $\pi$,
$\bbC_i, i \in[1,k]$ denotes the $i$-th chemical graph,
$m$ is the number of different kinds of environments,
and $\tau \in \bbR^m$ is an $m$-dimension real vector that represents the environment.
The observed property value corresponding to the $(k+1)$-tuple $(\bbC_1, \bbC_2, ..., \bbC_k, \tau)$,
i.e., $k$ chemical graphs $\bbC_1, \bbC_2, ..., \bbC_k$ under the environment $\tau$,
 is denoted by $a(\bbC_1, \bbC_2, ..., \bbC_k, \tau)$. 
A data set for the property $\pi$ is a set $D_\pi$ of  $(k+1)$-tuples $(\bbC_1, \bbC_2, ..., \bbC_k, \tau)$,
such that  $a(\bbC_1, \bbC_2, ..., \bbC_k, \tau)$ is available for every $(k+1)$-tuple. 
The purpose of Stage 1 is to collect data sets for the property $\pi$.

In Stage~2, let $f_i: \calG \to \bbR^{K_i}$ be a feature function defined for the $i$-th chemical graph $\bbC_i$, 
and $f_\tau: \bbR^m \to \bbR^{K_\tau}$ be a feature function that characterizes the environment $\tau$,
we define the feature function $f_\pi: D_\pi \to \bbR^{K_\pi}$ for the property $\pi$ to be the concatenation of the ones for each chemical graph and the one for the environment:
\[
f_\pi(\bbC_1, \bbC_2, ..., \bbC_k, \tau) \triangleq [ f_1(\bbC_1) \ f_2(\bbC_2) \  \cdots \ f_k(\bbC_k) \  f_\tau(\tau) ],
\]
where $K_\pi \triangleq \sum_{i=1}^k K_{i} + K_\tau$.
This feature function integrates the information from $k$ molecules and the environment.
Different choices of feature functions for each molecule and the environment may influence
the learning performance in Stage~3. 
We will have a brief discussion about the choices of the feature function for the environment
in the context of $\chi$-parameter in Section~\ref{sec:exp_phase1}.

Stage~3 is the same as \molinfer, i.e., it utilizes machine learning methods like linear regression or ANN 
to construct a prediction function $\eta$ between the feature vector $f_\pi(\bbC_1, \bbC_2, ..., \bbC_k, \tau)$ and the observed value
$a(\bbC_1, \bbC_2, ..., \bbC_k, \tau)$.

\subsection{Phase~2}\label{sec:formulation_phase2}

\begin{figure}[t!]
\begin{center} 
 \includegraphics[width=.95\columnwidth]{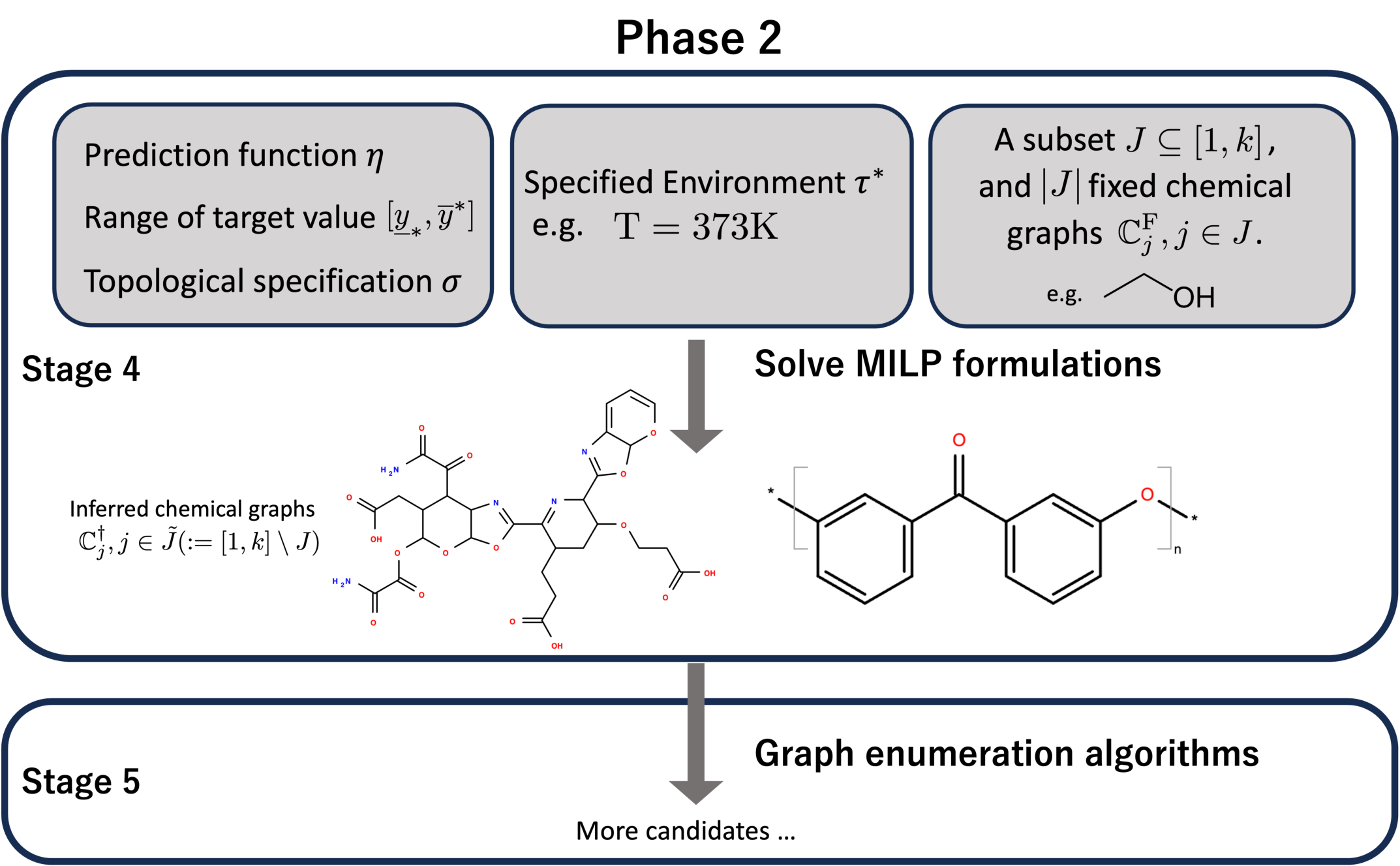}
\end{center}
\caption{An illustration of Phase 2 of the extended framework.
}
\label{fig:figure_2}  
\end{figure}

We illustrate Phase~2 of the extended framework in Figure~\ref{fig:figure_2}.

In Stage~4, 
besides the  topological specification $\sigma$ and a desired range $[\ylb, \yub]$ of the target value,
we also require a specified environment $\tau^*$, a subset $J \subseteq [1,k]$ ($J$ can be an empty set), 
and a set of chemical graphs $\bbC_j^\mathrm{F}, j \in J$.
This specified environment $\tau^*$ is introduced due to the practical consideration that
the experimental conditions are often fixed in real-world scenarios,
and the subset $J$ is included to account for
the possibility that certain molecules may be fixed
in the inverse phase.

Let $\tilde{J} := [1,k] \setminus J$,
we formulate an MILP formulation to infer the remaining $|\tilde{J}|$ chemical graphs $\bbC_j^\dagger, j \in \tilde{J}$
such that $\eta(f_\pi(\bbC_1^*, \bbC_2^*, ..., \bbC_k^*, \tau^*)) \in [\ylb, \yub]$, 
where $\bbC_j^*$ denotes the inferred chemical graph $\bbC_j^\dagger$ (resp., the fixed chemical graph $\bbC_j^\mathrm{F}$) 
if $j \in \tilde{J}$ (resp., $j \in J$).

In Stage~5,  we use some graph enumeration algorithms for each chemical graph $\bbC_j^\dagger$, $j \in \tilde{J}$ to obtain more candidates, as in \molinfer.



\section{Experimental Results}\label{sec:experiment}

To evaluate the effectiveness of our proposed model, 
we choose the property Flory-Huggins $\chi$-parameter which characterizes
the interaction between two molecules, solute $\bbC_1$ and solvent $\bbC_2$ 
to conduct numerical experiments.
Since it is well-known that the $\chi$-parameter value varies according to the temperature $T$, 
even for the same pair of solute-solvent pair\cite{Aoki:2023aa, Nistane:2022aa},
we consider the environment $\tau$ for the property $\chi$-parameter as a scalar representing the temperature $T$,
i.e., $\tau = (T)$, and thus one instance of $\chi$-parameter will be a triplet $(\bbC_1, \bbC_2, T)$.
Other environmental factors, such as pressure, are excluded due to their unavailability
in the data sets we collected.

We implemented the extended framework described in Section~\ref{sec:formulation} with a given range of $\chi$-parameter value and conducted experiments to evaluate the effectiveness.
The experiments were conducted on a PC with a Processor: Core i7-10700 (2.9GHz; 4.8 GHz at the maximum) and Memory: 16GB RAM DDR4.

\subsection{Experimental Results on Phase~1}\label{sec:exp_phase1}

We began by collecting two experimental $\chi$-parameter data sets that are
prepared by Aoki~et~al.~\cite{Aoki:2023aa} (\chiaoki) and 
Nistane~et~al.~\cite{Nistane:2022aa} (\chinistane), respectively. 
These two data sets were chosen for their experimental nature,
the relatively large number of available instances,
and the inclusion of temperature information for all solute-solvent pairs.
While the ``Flory-Huggins $\chi$ Database" from~\cite{pppdb} is also a widely-used data set in $\chi$-parameter studies, 
we opted to exclude it from our experiment 
due to the lack of temperature information for a significant number of solute-solvent pairs.

In addition to these two data sets, we also constructed a simulation-based $\chi$-parameter data set, \chijsol,
using the simulation software \jocta~\cite{JOCTA}.
In this data set, the $\chi$-parameter values between two molecules are calculated using the following formula:

\[
\chi = \chi_s + \frac{V}{R T} (\delta_A - \delta_B)^2,
\]
where $V$ represents the molar volume of the polymer-polymer/polymer-solvent system, 
$R = 8.314~\si{\joule / (\mole.\kelvin)}$ is the ideal gas constant, $T$ is the temperature in Kelvin ($\si{\kelvin}$), 
$\chi_s$ accounts for contributions besides enthalpy, and $\delta_A$ and $\delta_B$ are the Hildebrand solubility parameter of the two molecules.
For generating this data set,
we fixed the molar volume at $100~\si{\centi \metre^{3}/\mole}$,
and $\chi_s$ is set to be $0.34$ in the case of polymer-solvent pair and $0.0$ in the case of polymer-polymer pair.
Hildebrand solubility parameters for polymers were
calculated using either the Fedors method~\cite{Fedors:1974aa} or
the Krevelen method~\cite{Van-Krevelen:2009aa},
both are the so-called group-contribution methods implemented in \jocta.

\begin{figure}[t!]
\begin{center} 
 \includegraphics[width=.89\columnwidth]{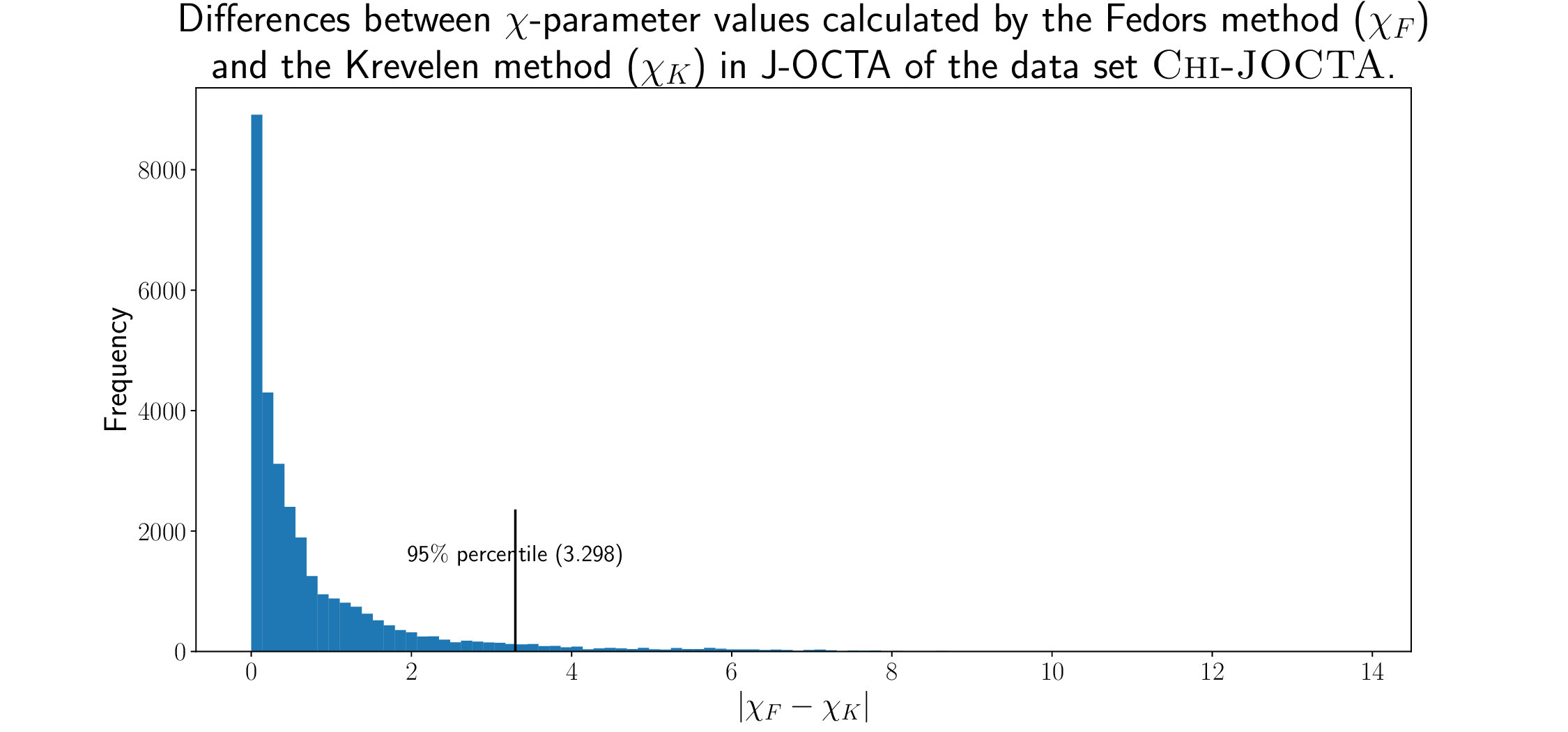}
\end{center}
\caption{A histogram of the differences between two methods (Fedors and Krevelen) available in \jocta\
for the data set \chijsol.
}
\label{fig:figure_chiFKdiff}  
\end{figure}

The two data sets \chiaoki\ and \chinistane\ consist of only polymer-solvent pairs, i.e., the solute $\bbC_1$ is always a polymer, 
and the solvent $\bbC_2$ is always a small molecule.
In contrast, the data set \chijsol\
was generated by first collecting 88 polymers and calculating the $\chi$-parameter values for all possible polymer-polymer pairs
at four distinct temperatures 
$T \in \{ 273 \si{\kelvin}, 298\si{\kelvin}, 323\si{\kelvin}, 348\si{\kelvin} \}$
using \jocta. In this process, the Fedors method was used to calculate the Hildebrand solubility parameters.
When generating this data set, we treated one polymer-polymer pair symmetrically, i.e., 
given a fixed temperature, two instances were created such that for one instance, one polymer was used as the solute and the other was the solvent,
and for the other instance, the roles of solute and solvent were reversed.
It is worth noting that the two available group-contribution methods (Fedors and Krevelen)
can produce different solubility parameter values.
Figure~\ref{fig:figure_chiFKdiff} presents a histogram of the differences between $\chi$-parameter values calculated
using these two methods for polymer-polymer pairs in the \chijsol\ data set.
Notably, the 95\% percentile of the differences is 3.298,
highlighting the variability in the results from the two approaches,
and also the difficulty of the task to accurately calculate $\chi$-parameter values.

For a data set $D_\chi$, we denote the set of all different solutes (resp., solvents) appearing in it by $D_{\chi, 1}$ (resp., $D_{\chi,2}$).
In the preprocessing procedure, we first convert all the polymers to their monomer representations that are defined in Section~\ref{sec:polymer}, and then remove some polymers such that a self-loop appears in the monomer form as in~\cite{Ido:2024aa} for each data set $D_\chi \in \{ $\chiaoki,  \chinistane, \chijsol$ \} $. 

We then calculated the proposed solute-solvent feature vectors $f_\chi(\bbC_1, \bbC_2, T)$ as defined in Section~\ref{sec:formulation_phase1}.
For the feature function $f_i$ used for each chemical graph $\bbC_i$, 
we choose the one consisting of only graph-theoretic descriptors defined 
in~\cite{Ido:2024aa} (resp., \cite{Zhu:2022ad}) in the case that $\bbC_i$ is a polymer (resp., a small molecule),
for $i=1, 2$, respectively.
See Appendix~\ref{sec:descriptor} for a complete list of descriptors used in the experiments.
For the environment $\tau = (T)$,
we tried three different kinds of feature functions as follows:
\begin{itemize}
\item[-] $f_{\tau, 1}(T) = (T) $;
\item[-] $f_{\tau,-1}(T) =  (\frac{1}{T}) $;
\item[-] $f_{\tau, 1, -1}(T) =  (T,  \frac{1}{T} )$.
\end{itemize}
These three simple feature functions for temperature are selected based on intuitive considerations. 
More complex transformations, such as $T^2$ or $\frac{1}{T^2}$, could introduce extreme value distributions
after standardization, making it less effective and harder to contribute to a prediction function.

Together with the one that neglects the environment $\tau$,
we obtained four kinds of feature vectors for a $\chi$-parameter instance  $(\bbC_1, \bbC_2, T)$ as follows:
\begin{itemize}
\item[-] $f_\chi (\bbC_1, \bbC_2, T) \triangleq [  f_1(\bbC_1) \ f_2(\bbC_2) ] $;
\item[-] $f_{\chi, 1}(\bbC_1, \bbC_2, T) \triangleq [ f_1(\bbC_1) \ f_2(\bbC_2) \ T ]$;
\item[-] $f_{\chi,-1}(\bbC_1, \bbC_2, T) \triangleq [ f_1(\bbC_1) \ f_2(\bbC_2) \ \frac{1}{T} ]$;
\item[-] $f_{\chi,1, -1}(\bbC_1, \bbC_2, T) \triangleq [ f_2(\bbC_1) \ f_2(\bbC_2) \ T \ \frac{1}{T} ]$.
\end{itemize}


\begin{table}[t!]\caption{Information on the data sets.}
\begin{center}\scalebox{0.99}{
\begin{tabular}{@{} c c c c c c c c c @{}} \hline
$D_\chi$ & $|D_\chi|$ & $\underline{n_1}, \overline{n_1}$ & $\underline{n_2}, \overline{n_2}$ & $\underline{a}, \overline{a}$ &  $|\mathcal{F}_1|$ & $K_1$ & $|\mathcal{F}_2|$ & $K_2$ \\ \hline
\chiaoki\cite{Aoki:2023aa} & 1190 & 2, 23 & 1, 16  & $-$2.24, 4.40  &  31 & 88 &  89 & 136 \\
\chinistane\cite{Nistane:2022aa} & 1581 & 2, 42 & 1, 16 & $-$0.55, 4.11 &   34 & 106 &  89 & 139 \\ 
\chijsol & 30624 & 2, 50 & 2, 50 & 0.0, 10.797 & 39 & 119 & 39 & 119 \\ \hline
\end{tabular}
}
\end{center}\label{table:phase1a}
\end{table}

We follow~\cite{Ido:2024aa} to first standardize the range of each descriptor and 
the range $\{ t \in \bbR \mid \underline{a} \leq t \leq \overline{a} \}$ of the property value $a(\bbC_1, \bbC_2, T)$.
Table~\ref{table:phase1a} summarizes 
some statistical information on the
data sets we prepared for $\chi$-parameter in Stage~1,
where the following notations are used:
\begin{itemize}
\item[-] $D_\chi$: the name of the data set;
\item[-] $|D_\chi|$: the total number of instances in $D_\chi$;
\item[-] $\underline{n_1}, \overline{n_1}$: the minimum and maximum values for the number of non-hydrogen atoms $n(\bbC_1)$ in solutes $\bbC_1$ within $D_{\chi, 1}$;
\item[-] $\underline{n_2}, \overline{n_2}$: the minimum and maximum values for the number of non-hydrogen atoms $n(\bbC_2)$ in solvents $\bbC_2$ within $D_{\chi, 2}$;
\item[-] $\underline{a}, \overline{a}$: the minimum and maximum values of $a(\bbC_1, \bbC_2, T)$ over the triplets $(\bbC_1, \bbC_2, T)$ in $D_\chi$;
\item[-] $|\mathcal{F}_1|$: the number of distinct non-isomorphic chemical rooted trees among all 2-fringe-trees in $D_{\chi, 1}$;
\item[-] $K_1$: the number of descriptors included in a feature vector $f(\bbC_1)$ for solutes $\bbC_1$;
\item[-] $|\mathcal{F}_2|$: the number of distinct non-isomorphic chemical rooted trees among all 2-fringe-trees in $D_{\chi, 2}$; and
\item[-] $K_2$: the number of descriptors included in a feature vector $f(\bbC_2)$ for solvents $\bbC_2$.
\end{itemize}

We used three different machine learning methods to 
construct a prediction function $\eta$ from the feature vector $f_\chi(\bbC_1, \bbC_2, T)$ 
to the observed value $a(\bbC_1, \bbC_2, T)$, 
namely ANN, 
multiple linear regression with reduced quadratic descriptors (R-MLR)~\cite{Zhu:2025aa}, 
and random forest (RF). 

For a given data set $D$,
let $x^i \triangleq f(\bbC_1^i, \bbC_2^i, T^i)$ and $a^i \triangleq a(\bbC_1^i, \bbC_2^i, T^i)$ for short,
where $(\bbC_1^i, \bbC_2^i, T^i) \in D$ represents an indexed triplet of instance.
To evaluate the learning performance of a prediction function $\eta$ on the data set $D$,
we define an error function
\[
\mathrm{Err}(\eta; D) \triangleq \sum_{(\bbC_1^i, \bbC_2^i, T^i) \in D} (a^i - \eta(x^i))^2,
\]
and the \emph{coefficient of determination} $\Rt(\eta, D)$ to be
\[
\Rt(\eta, D) \triangleq 1 - \frac{\mathrm{Err}(\eta; D)}{\sum_{(\bbC_1^i, \bbC_2^i, T^i) \in D} (a^i - \tilde{a})^2},
\]
 where $\tilde{a} \triangleq \frac{1}{|D|} \sum_{(\bbC_1^i, \bbC_2^i, T^i) \in D} a^i$  the average property value in the data set $D$.

\begin{table}[t!]\caption{Results of Stage~3 for the data set \chiaoki.}
\begin{center}\scalebox{0.99}{
\begin{tabular}{@{} c c c c @{}} \hline
 $f_\chi$ & ML & Train $\Rt$ & Test $\Rt$  \\ \hline
 & ANN & 0.922 & 0.725 \\
 $f_\chi(\bbC_1, \bbC_2)$ & R-MLR & 0.892 & 0.841 \\
& RF & 0.917 & 0.753 \\ \hline
& ANN & 0.922  & 0.725 \\
 $f_{\chi,1}(\bbC_1, \bbC_2, T)$ & R-MLR & 0.919 & *{\bf 0.891} \\
 & RF & 0.971 & 0.822 \\ \hline
 & ANN & 0.990 & {\bf 0.801} \\
 $f_{\chi,-1}(\bbC_1, \bbC_2, T)$ & R-MLR & 0.919 & 0.863 \\
 & RF & 0.973 & {\bf 0.822} \\ \hline
 & ANN & 0.990 & 0.767 \\
 $f_{\chi,1,-1}(\bbC_1, \bbC_2, T)$ & R-MLR & 0.921 & 0.879  \\
 & RF & 0.971  & 0.819 \\ \hline
\end{tabular}
}
\end{center}\label{table:phase1b_aoki}
\end{table}

\begin{table}[t!]\caption{Results of Stage~3 for the data set \chinistane.}
\begin{center}\scalebox{0.99}{
\begin{tabular}{@{} c c c c @{}} \hline
 $f_\chi$ & ML & Train $\Rt$ & Test $\Rt$  \\ \hline
 & ANN & 0.863 & 0.671 \\
$f_\chi(\bbC_1, \bbC_2)$ & R-MLR & 0.835 & 0.776 \\
 & RF & 0.882 & 0.731 \\ \hline
 & ANN & 0.971  & {\bf 0.684} \\
 $f_{\chi,1}(\bbC_1, \bbC_2, T)$ & R-MLR & 0.840 & {\bf 0.793} \\
 & RF & 0.968 & 0.793 \\ \hline
 & ANN & 0.950 & 0.655 \\
 $f_{\chi,-1}(\bbC_1, \bbC_2, T)$ & R-MLR & 0.824 & 0.768 \\
 & RF & 0.968 & 0.792 \\ \hline
 & ANN & 0.871 & 0.647 \\
$f_{\chi,1,-1}(\bbC_1, \bbC_2, T)$ & R-MLR & 0.860 & *{\bf 0.801}  \\
 & RF & 0.967  & 0.792 \\ \hline
\end{tabular}
}
\end{center}\label{table:phase1b_nistane}
\end{table}

\begin{table}[t!]\caption{Results of Stage~3 for the data set \chijsol.}
\begin{center}\scalebox{0.99}{
\begin{tabular}{@{} c c c c @{}} \hline
 $f_\chi$ & ML & Train $\Rt$ & Test $\Rt$  \\ \hline
 & ANN & 0.980 & 0.975  \\
$f_\chi(\bbC_1, \bbC_2)$ & R-MLR & 0.745 & 0.741  \\
 & RF & 0.987  &  0.977 \\ \hline
 & ANN & 0.993  &  0.990 \\
 $f_{\chi,1}(\bbC_1, \bbC_2, T)$ & R-MLR & 0.756 & 0.750  \\
 & RF & 0.998 & *{\bf 0.991} \\ \hline
 & ANN & 0.993  & {\bf 0.991}  \\
 $f_{\chi,-1}(\bbC_1, \bbC_2, T)$ & R-MLR & 0.787  & {\bf 0.782}  \\
 & RF & 0.998  & 0.991 \\ \hline
 & ANN & 0.993  & 0.991  \\
$f_{\chi,1,-1}(\bbC_1, \bbC_2, T)$ & R-MLR & 0.762 & 0.758  \\
 & RF & 0.998  & 0.991 \\ \hline
\end{tabular}
}
\end{center}\label{table:phase1b_jsol}
\end{table}

In Stage~3, we first conducted a preliminary experiment to decide the hyperparameter used for 
each machine learning method of ANN, R-MLR, and RF, and then conducted the experiments to 
evaluate the learning performance of each prediction function for each data set based on cross-validation. 
Please refer~\cite{Azam:2021ab, Zhu:2025aa} for the details about how we conducted these preliminary experiments.
For a given data set $D$, a \emph{cross-validation} procedure consists of five trials of constructing a prediction function as follows. 
The data set $D$ is first randomly divided into five subsets $D^{(k)}$, $k \in [1,5]$. For each trial $k \in [1,5]$, a prediction function $\eta^{(k)}$ is constructed using the hyperparameter determined in the preliminary experiment, with the subset $D \backslash D^{(k)}$ serving as the training data set, respectively.
We did ten times cross-validations for each data set and each kind of feature vector proposed in Section~\ref{sec:formulation_phase1}. 
The learning results are summarized in Tables~\ref{table:phase1b_aoki} to \ref{table:phase1b_jsol} 
for the data sets \chiaoki, \chinistane\ and \chijsol, respectively, where the following notations are used:
\begin{itemize}
\item[-] $f_\chi$: the feature vector for the triplet $(\bbC_1, \bbC_2, T)$ proposed in Section~\ref{sec:formulation_phase1};
\item[-] ML: the machine learning method used to construct a prediction function;
\item[-] Train/Test $\Rt$: the median of train/test $\Rt$ scores over all 50 trials in ten cross-validations; and
\item[-] the bold score indicates the highest $\Rt$ score achieved across the four different feature vectors for each learning method,
while the score marked with an asterisk denotes the overall highest $\Rt$ score among the three machine learning methods for each data set.
\end{itemize}


We can observe from Tables~\ref{table:phase1b_aoki} to~\ref{table:phase1b_jsol} that, 
in general, the feature function containing the information of the temperature 
enjoys a better learning performance than the simple one $f_\chi(\bbC_1, \bbC_2)$,
which matches our intuition.

As a comparison, Aoki et al.~\cite{Aoki:2023aa} attained an average $\Rt$ score of 0.834 on one-time cross-validation
 by using neural networks on the data set \chiaoki, 
 and Nistane et al.~\cite{Nistane:2022aa} attained a $\Rt$ score of 0.83 on a 10\% test set
  by utilizing Gaussian process regression (GPR) on the data set \chinistane. Although the specific way to evaluate the learning performance used in these works is different from ours, we demonstrate that our proposed feature vector can attain a competitively high learning performance.

\subsection{Experimental Results on Phase~2}\label{sec:exp_phase2}

\begin{figure}[t!]
\begin{center} 
 \includegraphics[width=.89\columnwidth]{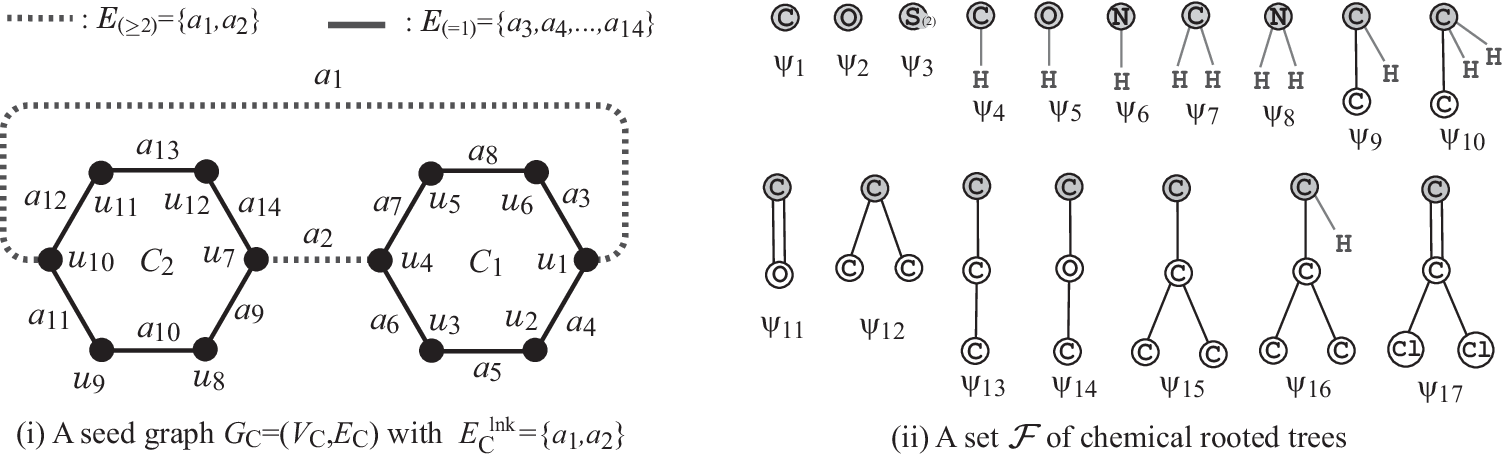}
\end{center}
\caption{(i) A seed graph $G_{\mathrm{C}}$ for $I_b$; (ii) A set $\mathcal{F}$ of chemical rooted trees.
The figure is adapted from~\cite{Ido:2024aa}.
}
\label{fig:seed_graph_b}  
\end{figure}

To execute Phase~2 to infer chemical graphs with desired $\chi$-parameter values, 
we used two instances $I_a$ and $I_b$ prepared by Ito~et~al.~\cite{Ido:2024aa}. 
Figures~\ref{fig:seed_graph_a} and~\ref{fig:seed_graph_b} 
illustrate the seed graphs $G_{\mathrm{C}}$ and the sets $\mathcal{F}$ of chemical rooted trees for $I_a$ and $I_b$, respectively. 
See Appendix~\ref{sec:test_instances} for a detailed description of the instances.


We executed Stage~4 for the two data sets $D \in \{ $\chiaoki, \chinistane $ \} $, 
with three different machine learning methods ANN, R-MLR, RF, respectively. 
Here the feature vector is selected as the one with the best median test $\Rt$ score attained 
for each learning method in Tables~\ref{table:phase1b_aoki} and \ref{table:phase1b_nistane}, and
the prediction functions used are the ones that attained the median test $\Rt$.
For the detailed MILP formulations that are formulated to simulate the prediction functions, 
we refer~\cite{Akutsu:2019aa} for ANN, and~\cite{Zhu:2025aa} for R-MLR, respectively.
The one for RF will appear in our future work,
while the formulation for decision tree is presented in~\cite{Tanaka:2021aa}.

For the inference of chemical graphs with desired $\chi$-parameter values,
we focus on the situation that the temperature $T$ and the solvent $\bbC_2$ are fixed, i.e., inferring the solute $\bbC_1$.
%
We selected carbon tetrachloride (resp., methyl acetate) as the solvent $\bbC_2^{\mathrm{F}}$ and both 373.5~\si{\kelvin} 
as the temperature during the experiment for the data set \chiaoki \ (resp., \chinistane). 
$\mathrm{CPLEX}$ 12.10 was used to solve the MILPs in Stage~4.

\begin{table}[t!]\caption{Results of Stages~4 and~5 of \chiaoki\ and \chinistane.}
\begin{center}\scalebox{0.69}{
\begin{tabular}{@{} c | c c c c c | c c c c c c | c c c @{}} \hline
No. & $D_\chi$ & $f_\chi$ & ML & inst. & $\underline{y}^*, \overline{y}^*$ & \#v & \#c & I-time & $n$ & $n^{\mathrm{int}}$ & $\eta$
& D-time & $\bbC$-LB & \#$\bbC$  \\ \hline
(a) & \chiaoki & $f_{\chi, -1}(\bbC_1, \bbC_2, T)$ & ANN & $I_a$ & 0.60, 0.70 & 10511 & 14017 & 518.289 & 48 & 27 & 0.624 & 0.114 & 8 & 8 \\
(b) &              & $                                        $ &           & $I_b$ & 1.00, 1.10 & 7062 & 8870 & 53.274 & 25 & 18 & 1.013 & 0.033 & 2 & 2 \\ \hline
(c) & \chiaoki & $f_{\chi, 1}(\bbC_1, \bbC_2, T)$ & R-MLR & $I_a$ & $-$0.10, 0.00 & 13177 & 17506 & 12.657 & 50 & 30 & -0.013 & 0.150 & 20 & 20 \\
(d) &              & $                                        $ &                  & $I_b$ & 1.50, 1.60 & 9728 & 12358 & 3.880 & 28 & 18 & 1.518 & 0.037 & 3 & 3 \\ \hline
(e) & \chiaoki & $f_{\chi, -1}(\bbC_1, \bbC_2, T)$ & RF & $I_a$ & 0.40, 0.50 & 62366 & 697356 & 33.213 & 50 & 28 & 0.476 & 0.115 & 8 & 8  \\
(f) &              & $                                        $ &       & $I_b$ & 0.60, 0.70 & 58917 & 692208 & 17.050 & 29 & 18 & 0.655 & 0.046 & 6 & 6  \\ \hline
(g) & \chinistane & $f_{\chi, 1}(\bbC_1, \bbC_2, T)$ & ANN & $I_a$ & 0.25, 0.35 & 12185 & 15198 & 37.684 & 50 & 30 & 0.303 & 0.194 & 32 & 32 \\
(h) &              & $                                        $ &              & $I_b$ & 0.40, 0.50 & 8061 & 9930 & 5.274 & 27 & 18 & 0.466 & 0.105 & 20 & 20 \\ \hline
(i) & \chinistane & $f_{\chi, 1, -1}(\bbC_1, \bbC_2, T)$ & R-MLR & $I_a$ & 0.60, 0.70 & 15035 & 18664 & 9.851 & 49 & 27 & 0.678 & 0.107 & 4 & 4 \\
(j) &              & $                                        $ &                  & $I_b$ & 3.20, 3.30 & 10913 & 13397 & 7.469 & 31 & 22 & 3.241 & 0.196 & 2706 & 100 \\ \hline
(k) & \chinistane & $f_{\chi, 1}(\bbC_1, \bbC_2, T)$ & RF & $I_a$ & 0.80, 0.90 & 80250 & 1001951 & 40.093 & 50 & 30 & 0.867 & 0.131 & 12 & 12 \\
(l) &              & $                                        $ &           & $I_b$ & 0.70, 0.80 & 76128 & 996684 & 25.653 & 28 & 18 & 0.742 & 0.035 & 2 & 2 \\ \hline
\end{tabular}
}
\end{center}\label{table:phase2}
\end{table}

The computational results of the experiment in Stage 4 are shown in
Table~\ref{table:phase2},
where the following notations are used:
\begin{itemize}
\item[-] No.: numbering of the MILP instance;
\item[-] $D_\chi$: name of the data set;
\item[-] $f_\chi$: the feature vector for the triplet $(\bbC_1, \bbC_2, T)$ proposed in Section~\ref{sec:formulation_phase1};
\item[-] ML: the machine learning method used to construct a prediction function;
\item[-] inst.: instance $I_a$ or $I_b$;
\item[-] $\underline{y}^*, \overline{y}^*$: range $[\underline{y}^*, \overline{y}^*]$ on the value $a(\bbC_1, \bbC_2, T)$ of a polymer $\bbC_1$ to be inferred;
\item[-] \#v (resp., \#c): the number of variables (resp., constraints) in the MILP; 
\item[-] I-time: the time (in seconds) to solve the MILP; 
\item[-] $n$: the number $n(\bbC_1^\dagger)$ of non-hydrogen atoms of the inferred polymer in the monomer representation $\bbC_1^\dagger$; 
\item[-] $n^{\mathrm{int}}$: the number $n^{\mathrm{int}}(\bbC_1^\dagger)$ of interior-vertices of the inferred polymer in the monomer representation $\bbC_1^\dagger$; and 
\item[-] $\eta$: the predicted $\chi$-parameter value $\eta(f_\chi(\bbC_1^\dagger, \bbC_2^{\mathrm{F}}, T))$ of the inferred polymer $\bbC_1^\dagger$. 
\end{itemize}

\begin{figure}[t!]
\begin{center} 
 \includegraphics[width=.99\columnwidth]{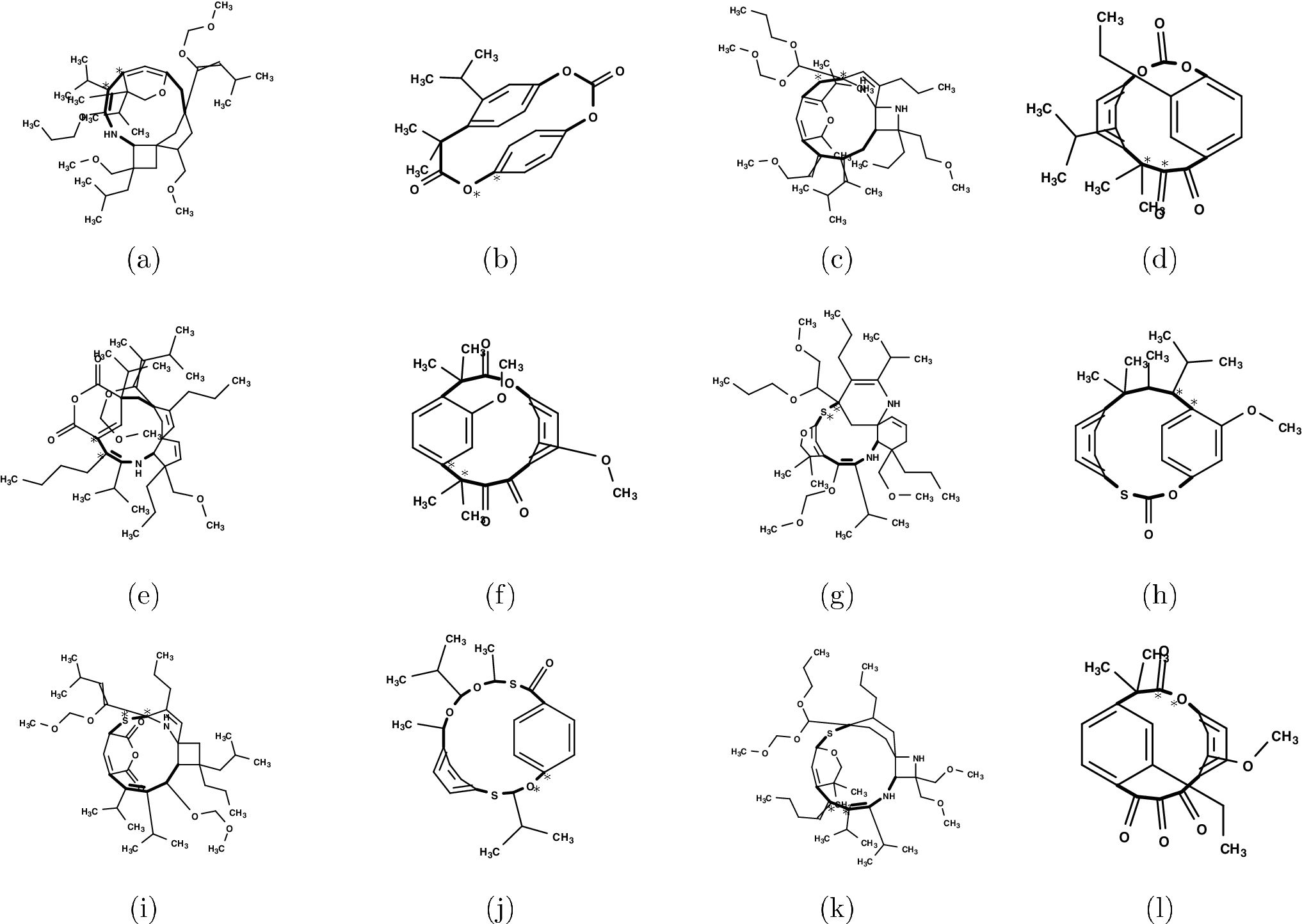}
\end{center}
\caption{Illustration of the solute polymers inferred in Stage~4. The link-edges are shown with thick lines, and the two connecting-vertices are marked with asterisks. (a)-(l) correspond to the first column No. in Table~\ref{table:phase2}, respectively.
}
\label{fig:milp_results}  
\end{figure}

From Table~\ref{table:phase2}, we observe that all 12 instances can be solved within 520 seconds, with 28 to 50 non-hydrogen atoms
in their monomer presentation,
and the predicted $\chi$-parameter values $\eta(f_\chi(\bbC_1^\dagger, \bbC_2^{\mathrm{F}}, T))$ 
all fall inside the range $[\underline{y}^*, \overline{y}^*]$ that we specified.
The inferred molecules are illustrated in Figure~\ref{fig:milp_results}.

Finally, we executed Stage~5 to generate more target chemical graphs $\bbC^*$, i.e., the chemical isomers of $\bbC^*$.
Utilizing the algorithm proposed by Zhu~et~al.~\cite{Zhu:2022ad},
we generated the chemical isomers for each target chemical graph $\bbC^\dagger$ inferred in Stage~4.
If the total number of chemical isomers exceeded 100, we limited the generation to the first 100 isomers.
Additionally, the algorithm can also provide an evaluation of a lower bound 
on the total number of possible chemical isomers of $\C^\dagger$.
The computational results are summarized in 
Table~\ref{table:phase2}, where the following notations are used:
\begin{itemize}
\item[-] D-time: the execution time (in seconds) of the algorithm in Stage 5;
\item[-] $\bbC$-LB: a lower bound on the number of all chemical isomers $\bbC^*_1$ of $\bbC_1^\dagger$; and
\item[-] \#$\bbC$: the number of generated chemical isomers $\bbC^*_1$ of $\bbC_1^\dagger$. 
\end{itemize}

It can be observed that the algorithm can find chemical isomers efficiently, 
in less than 0.2 seconds for all cases.

\subsection{Comparison with the Simulation Software \jocta}
In this subsection, we present experimental results of inferring chemical graphs based on the data set \chijsol\
to demonstrate that the inferred polymer-polymer pairs are of good quality, by comparing the $\chi$-parameter
values inferred by prediction functions to those computed using the simulation software \jocta~\cite{JOCTA}.

For this purpose, we prepared two instances $I_{\mathrm{c1}}$ and $I_{\mathrm{c2}}$ which are similar to the polymers in our collected data sets.
The two seed graphs for $I_{\mathrm{c1}}$ and $I_{\mathrm{c2}}$ are illustrated in Figure~\ref{fig:figure_c1c2}(a) and (b), respectively, with the detailed descriptions provided in Appendix C.
Instance $I_{\mathrm{c1}}$ is designed to represent a relatively simple polymer structure, whereas
instance $I_{\mathrm{c2}}$ is introduced for the situation that a benzene ring is included.
Although RF achieved the best learning performance in Table~\ref{table:phase1b_jsol},
we selected ANN as the learning method for constructing the prediction function.
This choice was motivated by the strong learning performance obtained by ANN and the significantly reduced
number of variables and constraints in the MILP compared to RF,
which can be observed in Table~\ref{table:phase2}.
For each instance, we selected four polymers 
(polyethylene (PE), polyethylene glycol (PEG), polystyrene sulfonates (PSS), and polyphthalamide (PPA)) 
as the fixed solvent $\bbC_2^{\mathrm{F}}$.
Additionally, we defined nine ranges of target $\chi$-parameter values ($[0.20, 0.70]$, $[0.40, 0.90]$, 
$[0.70, 1.20]$, $[0.90, 1.40]$, $[1.20, 1.70]$, $[1.40, 1.90]$, $[1.70, 2.20]$, $[1.90, 2.40]$, and $[2.20, 2.70]$)
and four temperatures ($273 \si{\kelvin}, 298\si{\kelvin}, 323\si{\kelvin},$ and $348\si{\kelvin}$)
to conduct the experiments, resulting in a total of 288 MILPs.

\begin{figure}[t!]
\begin{center} 
 \includegraphics[width=.89\columnwidth]{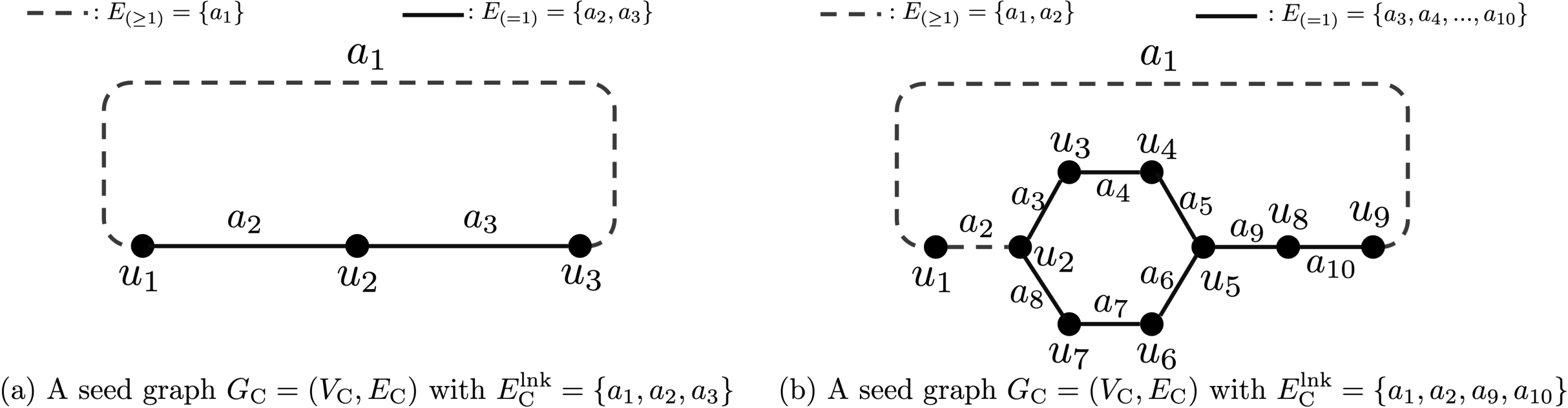}
\end{center}
\caption{Seed graphs for instances (a) $I_{\mathrm{c1}}$ and (b) $I_{\mathrm{c2}}$, respectively.
}
\label{fig:figure_c1c2}  
\end{figure} 

For these experiments, ANN was used
as the machine learning method to construct the prediction function
and $\mathrm{CPLEX}$ 12.10 was used to solve the MILPs,
with a time limit of 3600 seconds for each MILP. 
Once a polymer was inferred from MILP,
we utilized \jocta\ to compute the $\chi$-parameter value between the polymer as the solute and the corresponding solvent
 using the same way for generating the data set \chijsol.

\begin{figure}[t!]
\begin{center} 
 \includegraphics[width=.89\columnwidth]{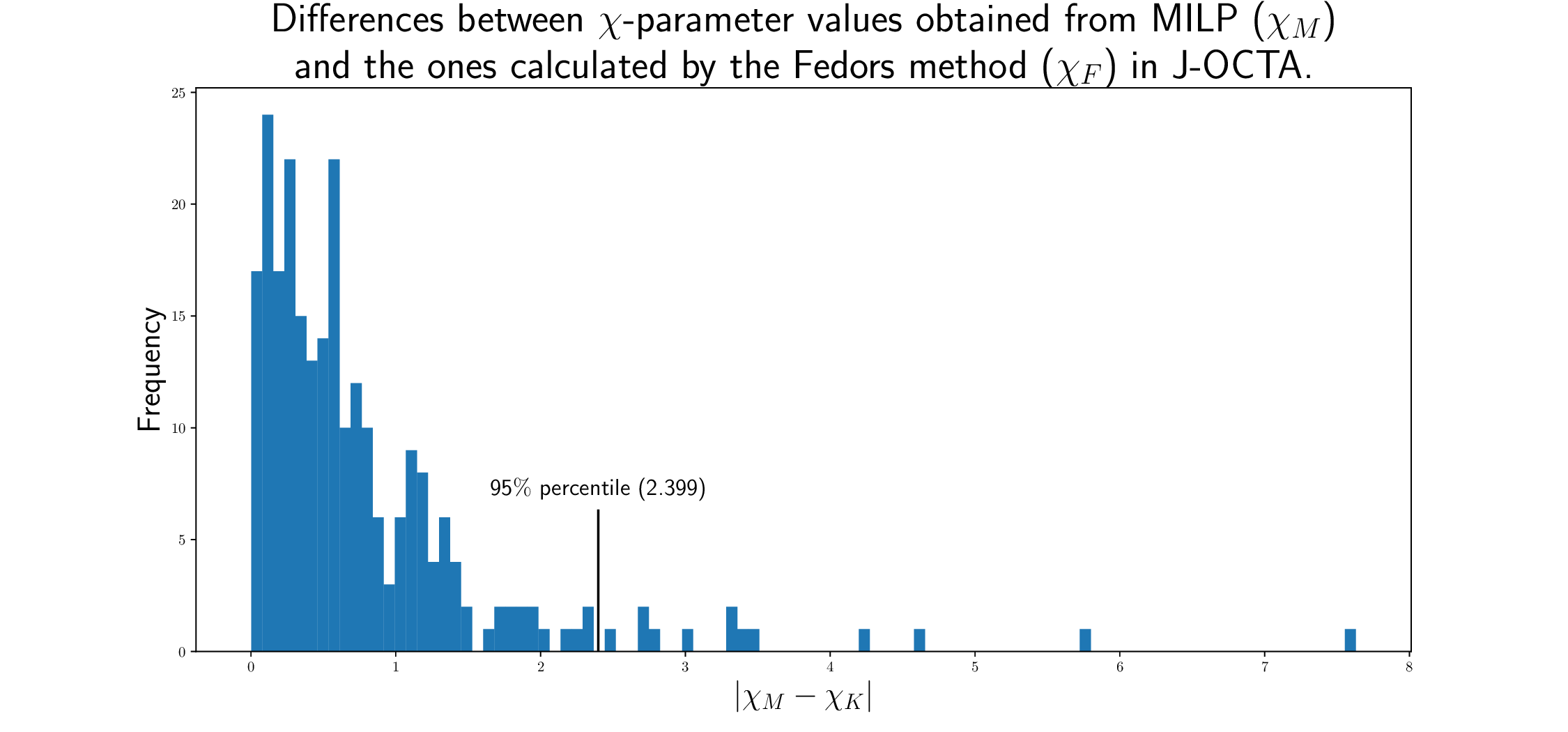}
\end{center}
\caption{A histogram of the differences between the Fedors method in \jocta\ and the value obtained from MILP.
}
\label{fig:figure_chiMFdiff}  
\end{figure}

Out of the 288 MILPs,
251 (87.2\%) were successfully solved within the time limit. 
We then computed the differences between the $\chi$-parameter values obtained from the MILPs
and those calculated using \jocta\ via the Fedors method.
Figure~\ref{fig:figure_chiMFdiff} presents a histogram of these differences,
showing that 95\% of the differences are less than 2.399.
Notably, the 95\% percentile for the differences between the MILP results and the Fedors method
is smaller than that for the differences between the Fedors method and Krevelen method, suggesting that
the differences are acceptable given that both methods are considered valid for predicting $\chi$-parameter values.
Since accurate calculation of $\chi$-parameter values is inherently challenging,
as evidenced in Figure~\ref{fig:figure_chiFKdiff},
where discrepancies between different methods are apparent,
we content ourselves that the polymers inferred by our MILP approach
are of high quality,
based on the relatively close agreement between the values obtained from MILP
and those calculated by \jocta.

\begin{figure}[t!]
\centering
\begin{minipage}{.45\textwidth}
\begin{center}
 \includegraphics[width=.45\columnwidth]{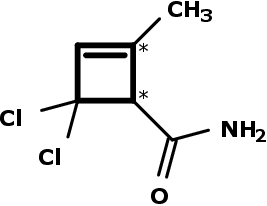}
\end{center}
\subcaption{}
\end{minipage}
\begin{minipage}{.45\textwidth}
\begin{center}
 \includegraphics[width=.45\columnwidth]{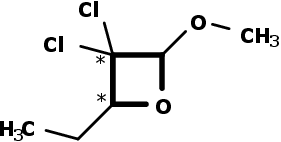}
\end{center}
\subcaption{}
\end{minipage}
\begin{minipage}{.45\textwidth}
\begin{center}
 \includegraphics[width=.45\columnwidth]{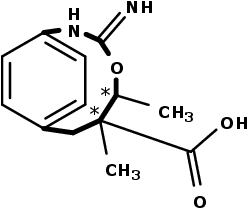}
\end{center}
\subcaption{}
\end{minipage}
\begin{minipage}{.45\textwidth}
\begin{center}
 \includegraphics[width=.45\columnwidth]{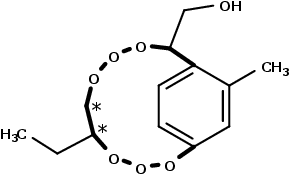}
\end{center}
\subcaption{}
\end{minipage}
\caption{Illustrations of the solute polymers inferred in Stage~4 using the data set \chijsol.
The link-edges are shown with thick lines, and the two connecting-vertices are marked with asterisks.
}
\label{fig:milp_results_JOCTA}  
\end{figure}

We illustrate some solutions whose $\chi$-parameter value obtained from MILP is close to the one calculated 
by \jocta\ using the Fedors method in Figure~\ref{fig:milp_results_JOCTA}. 
The detailed results for them are summarized in Table~\ref{table:milp_JOCTA}, where we denote the following:
\begin{itemize}
\item[-] No.: numbering of the MILP instance;
\item[-] $f_\chi$: the feature vector for the triplet $(\bbC_1, \bbC_2, T)$ proposed in Section~\ref{sec:formulation_phase1};
\item[-] ML: the machine learning method used to construct a prediction function;
\item[-] inst.: instance $I_{\mathrm{c1}}$ or $I_{\mathrm{c2}}$;
\item[-] $\underline{y}^*, \overline{y}^*$: range $[\underline{y}^*, \overline{y}^*]$ on the value $a(\bbC_1, \bbC_2, T)$ of a polymer $\bbC_1$ to be inferred;
\item[-] $\bbC_2^{\mathrm{F}}$: the selected solvent; 
\item[-] $T$: the selected temperature; 
\item[-] I-time: the time (in seconds) to solve the MILP; 
\item[-] $\chi_M$: the predicted $\chi$-parameter value $\eta(f_\chi(\bbC_1^\dagger, \bbC_2^{\mathrm{F}}, T))$ of the inferred polymer $\bbC_1^\dagger$; 
\item[-] $\chi_F$: the $\chi$-parameter value calculated between $\bbC_1^\dagger$ and $\bbC_2^{\mathrm{F}}$
under the temperature $T$ using the Fedors method implemented in \jocta; and
\item[-] $|\chi_M - \chi_F|$: the absolute difference between $\chi_M$ and $\chi_F$.
\end{itemize}


\begin{table}[t!]\caption{Selected results of Stage~4 for the data set \chijsol}
\begin{center}\scalebox{0.85}{
\begin{tabular}{|c|cccccc|c|cc|c|}
\hline
No. & $f_\chi$ & ML & inst. & $\underline{y}^*, \overline{y}^*$ &  $\bbC_2^{\mathrm{F}}$ & $T$ & I-time & $\chi_M$ & $\chi_F$ & $|\chi_M - \chi_F|$ \\ \hline
(a) & $f_{\chi, -1}(\bbC_1, \bbC_2, T)$ & ANN & $I_{c_1}$ & $0.40, 0.90$ & PPA & 298 $\si{\kelvin}$  & 14.723 & 0.790285 & 0.813052 & 0.022766 \\ \hline 
(b) & $f_{\chi, -1}(\bbC_1, \bbC_2, T)$ & ANN & $I_{c_1}$ & $0.90, 1.40$ & PSS & 273 $\si{\kelvin}$  & 13.305 & 1.149399 & 1.147394 & 0.002005 \\ \hline 
(c) & $f_{\chi, -1}(\bbC_1, \bbC_2, T)$ & ANN & $I_{c_2}$ & $0.20, 0.70$ & PE   & 348 $\si{\kelvin}$ & 34.422  & 0.644275 & 0.723993  & 0.079719 \\ \hline
(d) & $f_{\chi, -1}(\bbC_1, \bbC_2, T)$ & ANN & $I_{c_2}$ & $0.70, 1.20$ & PEG   & 298 $\si{\kelvin}$ & 65.631  & 0.972358 & 0.963446  & 0.008912 \\ \hline
\end{tabular}
\label{table:milp_JOCTA}
}
\end{center}
\end{table}

As far as we know, there is no method that conducts
 inference of molecules or molecule pairs
with specific $\chi$-parameter values,
therefore we believe that our proposed method 
represents a significant advancement in this area and related fields.

\section{Concluding Remarks}\label{sec:conclude}

We proposed an extended inverse QSAR/QSPR model
to infer molecules with desired properties under specified environments in this paper.
Building upon the existing framework \molinfer,
which integrates MILP and machine learning,
we extended its capabilities by integrating the information for each molecule and the environment into a unified feature vector,
formulating the inverse problem as an MILP.

To evaluate the computational effectiveness of our proposed model,
we particularly focused on the property Flory-Huggins $\chi$-parameter,
which characterizes the thermodynamic interaction between the solute and the solvent and is temperature-dependent.
We designed feature vectors that incorporate solute and solvent information, 
with several different ways to integrate the temperature information.
Utilizing machine learning techniques such as
artificial neural networks (ANN),  multiple linear regression with reduced quadratic descriptors (R-MLR), 
and random forest (RF), we conducted numerical experiments to demonstrate
that our proposed feature vectors can achieve competitively high $\Rt$ scores.

Furthermore, we addressed the task of designing solutes
for some desired $\chi$-parameter values when paired with a given solvent
under specified temperature conditions.
By formulating this task as an MILP,
our experimental results showed that polymers with up to 50 non-hydrogen atoms
in their monomer representations could be identified within reasonable computation times.
Additionally, using an efficient graph enumeration algorithm, 
more candidates can be generated rapidly by enumerating chemical isomers of the inferred polymers.
We also validated the quality of the inferred molecules using a simulation-based data set using the simulation software \jocta,
demonstrating that $\chi$-parameter values calculated using \jocta\ software were relatively closely aligned
with the ones obtained from our method. This underscores the accuracy and practical utility of our approach.
We believe that our proposed framework represents a significant advancement in this field,
with the potential to drive innovation in molecular design and related applications.


Future research could focus on designing more sophisticated feature functions 
to better capture environmental complexities and their interactions with molecular systems,
further enhancing the framework's predictive and inferential performance.
Also, we are interested in applying the framework
to the problem of deciding whether chemical reactions occur or not among several molecules and specific environments,
which is of great importance.

\section*{Acknowledgement}
This work is partially supported by JSPS KAKENHI Grant Numbers JP22H00532 and JP22KJ1979. 
We also express our gratitude to JSOL Corporation for generously providing part of the data sets used in this research.


\bibliographystyle{abbrv}
\bibliography{./chemgraph.bib}

\appendix

\newpage
\section*{Appendix}

\section{A Full Description of Descriptors}\label{sec:descriptor}

Our definition of feature function is analogous to the one  by 
 Zhu~et~al.~\cite{Zhu:2022ad}. When the molecule is a polymer, we follow Ido~et~al.~\cite{Ido:2024aa}
 to add some additional descriptors.

Associated with the two functions 
$\alpha$ and $\beta$ in a chemical graph $\Co=(H,\alpha,\beta)$,
we introduce   functions  
 $\ac: V(E)\to (\Lambda\setminus\{\ttH\})\times (\Lambda\setminus\{\ttH\})\times [1,3]$, 
 $\cs: V(E)\to (\Lambda\setminus\{\ttH\})\times [1,6]$ and
$\ec: V(E)\to ((\Lambda\setminus\{\ttH\})\times [1,6])
\times ((\Lambda\setminus\{\ttH\})\times [1,6])\times [1,3]$
in the following. 

 To represent  a feature of the exterior  of  $\Co$, 
  a  chemical rooted tree in $\mathcal{T}(\Co)$ is
  called a {\em fringe-configuration} of $\Co$. 

We also represent leaf-edges in the exterior of $\Co$.
For a leaf-edge $uv\in E(\anC)$ with $\deg_{\anC}(u)=1$, we define
the {\em adjacency-configuration} of $e$ to be an ordered tuple
$(\alpha(u),\alpha(v),\beta(uv))$. 
Define 
\[ \Gac^\lf\triangleq \{(\ta,\tb,m)\mid \ta,\tb\in\Lambda, 
m\in[1,\min\{\val(\ta),\val(\tb)\}]\} \]
as a set of possible adjacency-configurations for leaf-edges. 

To  represent a feature of an interior-vertex $v\in V^\inte(\Co)$ such that
$\alpha(v)=\ta$  and  $\deg_{\anC}(v)=d$
(i.e., the number of non-hydrogen atoms adjacent to $v$ is $d$) 
   in a chemical   graph  $\Co=(H,\alpha,\beta)$,
 we use  a pair $(\ta, d)\in (\Lambda\setminus\{{\tt H}\})\times [1,4]$,
 which we call the {\em chemical symbol} $\cs(v)$ of the vertex $v$.
 We treat $(\ta, d)$ as a single symbol $\ta d$,  and  
define $\Ldg$   to be  the set of all chemical symbols
$\mu=\ta d\in  (\Lambda\setminus\{{\tt H}\})\times [1,4]$.  

We define a method for featuring interior-edges  as follows.
Let $e=uv\in E^\inte(\Co)$  be 
 an interior-edge $e=uv\in E^\inte(\Co)$ 
 such that $\alpha(u)=\ta$, $\alpha(v)=\tb$ and $\beta(e)=m$ 
   in a chemical graph  $\Co=(H,\alpha,\beta)$.
To feature this edge $e$, 
 we use a tuple $(\ta,\tb,m)\in (\Lambda\setminus\{{\tt H}\})
    \times (\Lambda\setminus\{{\tt H}\})\times [1,3]$,
 which we call the {\em adjacency-configuration} $\ac(e)$ of the edge $e$. 
 We introduce a total order $<$ over the elements in $\Lambda$
 to distinguish  between $(\ta,\tb, m)$ and $(\tb,\ta, m)$ 
 $(\ta\neq \tb)$ notationally.
 For a tuple  $\nu=(\ta,\tb, m)$,
 let $\overline{\nu}$ denote the tuple $(\tb,\ta, m)$.  

Let $e=uv\in E^\inte(\Co)$  be 
an  interior-edge $e=uv\in E^\inte(\Co)$ 
 such that $\cs(u)=\mu$, $\cs(v)=\mu'$ and $\beta(e)=m$ 
   in a chemical  graph  $\Co=(H,\alpha,\beta)$.
To feature this edge $e$, 
 we use a tuple $(\mu,\mu',m)\in \Ldg\times \Ldg\times [1,3]$, 
 which we call  the {\em edge-configuration} $\ec(e)$ of the edge $e$. 
 We introduce a total order $<$ over the elements in $\Ldg$
 to distinguish between $(\mu,\mu', m)$ and $(\mu', \mu, m)$ 
 $(\mu \neq \mu')$ notationally. 
 For a tuple  $\gamma=(\mu,\mu', m)$,
 let $\overline{\gamma}$ denote the tuple $(\mu', \mu, m)$. 
   
Let $\pi$ be a chemical property for which we will construct
a prediction function $\eta$ from a feature
vector $f(\C)$ of a chemical graph $\C$ 
to a predicted value $y\in \mathbb{R}$
for the  chemical property of $\C$.

We first choose a set $\Lambda$ of chemical elements
 and then collect a data set  $D_{\pi}$ of
  chemical compounds  $C$ whose 
  chemical elements belong to $\Lambda$,
  where we regard  $D_{\pi}$ as a set of chemical graphs $\C$
  that represent the chemical compounds $C$  in  $D_{\pi}$.
To define the interior/exterior of 
chemical graphs  $\C\in D_{\pi}$,
we  next choose a branch-parameter ${\rho}$, where
 we recommend ${\rho}=2$.  
 
Let $\Lambda^\inte(D_\pi)\subseteq \Lambda$  
(resp., 
$\Lambda^\ex(D_\pi)\subseteq \Lambda$)
denote the set  of chemical elements  used in
the set $V^\inte(\C)$ of interior-vertices
(resp., the set $V^\ex(\C)$ of  exterior-vertices) of $\C$
 over all chemical graphs $\C\in D_\pi$, 
and $\Gamma^\inte(D_\pi)$
(resp., $\Gamma^\lnk(D_\pi)$) 
denote the set of edge-configurations used in
the set $E^\inte(\C)$  of interior-edges
(resp., the set $\Elnk(\C)$ of linked-edges) in $\C$
 over all chemical graphs $\C\in D_\pi$. 
Let $\mathcal{F}(D_\pi)$ denote the set of
chemical rooted trees $\psi$  
r-isomorphic to a chemical rooted tree in $\mathcal{T}(\C)$
  over all chemical graphs $\C\in D_\pi$,
  where possibly a chemical rooted tree $\psi\in \mathcal{F}(D_\pi)$
  consists of a single chemical element $\ta\in \Lambda\setminus \{{\tt H}\}$.
  
We define an integer encoding of a finite set $A$ of elements
to be a bijection $\sigma: A \to [1, |A|]$, 
where we denote by $[A]$   the set $[1, |A|]$ of integers.
Introduce  an integer coding of each of the   sets 
$\Lambda^\inte(D_\pi)$, $\Lambda^\ex(D_\pi)$, 
$\Gamma^\inte(D_\pi)$ and $\mathcal{F}(D_\pi)$. 
Let $[\ta]^\inte$  
(resp., $[\ta]^\ex$)  denote   
the coded integer of  an element $\ta\in \Lambda^\inte(D_\pi)$
(resp., $\ta\in \Lambda^\ex(D_\pi)$),  
$[\gamma]$   denote  
the coded integer of  an element $\gamma$ in $\Gamma^\inte(D_\pi)$
and 
$[\psi]$   denote  an element $\psi$ in $\mathcal{F}(D_\pi)$. 
 
We assume that a chemical graph $\C$
 treated in this paper satisfies  $\deg_{\anC}(v)\leq 4$
in the hydrogen-suppressed graph $\anC$.
 
In our model, we  use an integer 
  $\mathrm{mass}^*(\ta)=\lfloor 10\cdot \mathrm{mass}(\ta)\rfloor$, 
 for each $\ta\in \Lambda$.
 
  We define the {\em feature vector} $f(\C)$ 
  of a molecule $\C=(H,\alpha,\beta)\in D_{\pi}$ 
  to be a vector that consists of the following  
non-negative integer descriptors $\dcp_i(\C)$, $i\in [1,K]$, where 
$K=14+ |\Lambda^\inte(D_\pi)|+|\Lambda^\ex(D_\pi)|
+|\Gamma^\inte(D_\pi)|+|\Gamma^\lnk(D_\pi)|+|\Ldg|
+|\mathcal{F}(D_\pi)|+|\Gac^\lf|$. 
Notice that some descriptors are used for the case of polymers only.


\begin{enumerate}  
\item   
$\dcp_1(\C)$: the number  $|V(H)|-|\VH|$ of non-hydrogen atoms  in  $\C$.  
 
\item 
$\dcp_2(\C)$:  the number $|V^\inte(\C)|$ of interior-vertices in  $\C$.
  
\item 
$\dcp_3(\C)$:  the number $|\Elnk(\C)|$ of link-edges in  $\C$.
This descriptor is only for the case of polymers.

\item 
$\dcp_4 (\C)$: 
the average $\overline{\mathrm{ms}}(\C)$ of mass$^*$ 
over all atoms in $\C$; \\
 i.e., $\overline{\mathrm{ms}}(\C)\triangleq 
 \frac{1}{|V(H)|}\sum_{v\in V(H)}\mathrm{mass}^*(\alpha(v))$. 

\item 
$\dcp_i(\C)$,  $i=4+d,   d\in [1,4]$: 
the number $\dg_d^{\oH} (\C)$  of non-hydrogen vertices $v\in V(H)\setminus \VH$
 of degree $\deg_{\anC}(v)=d$
 in the hydrogen-suppressed chemical graph $\anC$.  
 
\item 
$\dcp_i(\C)$,  $i=8+d,   d\in [1,4]$: 
the number $\dg_d^\inte(\C)$
 of interior-vertices of interior-degree  $\deg_{\C^\inte}(v)=d$
  in the interior $\C^\inte=(V^\inte(\C),E^\inte(\C))$ of  $\C$. 
  
   
\item $\dcp_i(\C)$, $i=12+m$,  $m\in[2,3]$: 
the number $\bd_m^\inte(\C)$
 of  interior-edges with bond multiplicity $m$ in  $\C$; 
 i.e., $\bd_m^\inte(\C)\triangleq \{e\in E^\inte(\C)\mid \beta(e)=m\}$.

\item $\dcp_i(\C)$, $i=14+[\ta]^\inte$, 
 $\ta\in \Lambda^\inte(D_\pi)$: 
 the frequency $\na_\ta^\inte(\C)=|V_\ta(\C)\cap V^\inte(\C) |$ 
 of chemical element $\ta$ in
 the set $V^\inte(\C)$ of  interior-vertices in  $\C$. 
 
\item $\dcp_i(\C)$, 
$i=14+|\Lambda^\inte(D_\pi)|+[\ta]^\ex$, 
 $\ta\in \Lambda^\ex(D_\pi)$: 
 the frequency $\na_\ta^\ex(\C)=|V_\ta(\C)\cap V^\ex(\C) |$
  of chemical element $\ta$ in
 the set $V^\ex(\C)$ of  exterior-vertices in  $\C$. 
 
\item $\dcp_i(\C)$, 
$i=14+|\Lambda^\inte(D_\pi)|+|\Lambda^\ex(D_\pi)|+ [\gamma]$, 
$\gamma \in \Gamma^\inte(D_\pi)$: 
the frequency $\ec_{\gamma} (\C)$ of edge-configuration $\gamma$
in the set $E^\inte(\C)$ of interior-edges   in  $\C$. 

\item $\dcp_i(\C)$, 
$i=14+|\Lambda^\inte(D_\pi)|+|\Lambda^\ex(D_\pi)|+ |\Gamma^\inte(D_\pi)|
+ [\gamma]$, 
$\gamma \in \Gamma^\lnk(D_\pi)$: 
the frequency $\ec_{\gamma} (\C)$ of edge-configuration $\gamma$
in the set $\Elnk(\C)$ of link-edges   in  $\C$. 
This descriptor is only for the case of polymers.

\item $\dcp_i(\C)$, 
$i=14+|\Lambda^\inte(D_\pi)|+|\Lambda^\ex(D_\pi)|+ |\Gamma^\inte(D_\pi)|
+ [\mu]$, 
$\mu\in \Ldg^\inte$: 
the frequency of chemical symbols $\mu=\alpha(u)\deg_{\anC}(u)$ 
 of connecting-vertices $u$   in  $\C$.
 This descriptor is only for the case of polymers.

\item $\dcp_i(\C)$, 
$i= 14+|\Lambda^\inte(D_\pi)|+|\Lambda^\ex(D_\pi)|
+|\Gamma^\inte(D_\pi)|+|\Gamma^\lnk(D_\pi)|+|\Ldg|+ [\psi]$,  
 $\psi \in \mathcal{F}(D_\pi)$: 
the frequency $\fc_{\psi}(\C)$ of fringe-configuration $\psi $
in the set of ${\rho}$-fringe-trees in  $\C$. 

\item $\dcp_i(\C)$, 
$i= 14+|\Lambda^\inte(D_\pi)|+|\Lambda^\ex(D_\pi)|
+ |\Gamma^\inte(D_\pi)|+|\Gamma^\lnk(D_\pi)|+|\Ldg|
+|\mathcal{F}(D_\pi)|+ [\nu]$,  
 $\nu \in \Gac^\lf$: 
the frequency $\ac_{\nu}^\lf(\C)$ of adjacency-configuration $\nu$
in the set of leaf-edges in  $\anC$. 
\end{enumerate}

\section{Specifying Target Chemical Graphs}\label{sec:specification}

Our definition of the topological specification is analogous to the one  by 
 Zhu~et~al.~\cite{Zhu:2022ad}.
 Here we review the one particularly modified for polymers proposed by~Ido~et~al.~\cite{Ido:2024aa}.

\subsection*{Seed Graph}

A  {\em seed graph} for a polymer  is defined
to be a graph $\GC=(\VC,\EC)$  with a specified edge subset $\EC^\lnk$
 such that 
the edge set $\EC$ consists of four sets 
$\Et$, $\Ew$, $\Ez$ and $\Eew$, 
where each of them can be empty, and
 $\EC^\lnk$ is a circular  set in $\GC$ such that 
  $\emptyset\neq \EC^\lnk\subseteq \Et\cup \Ew\cup \Eew$ (only for polymer). 
Figure~5(a) 
 illustrates an example of a seed graph,
where $\VC=\{u_1,u_2,\ldots,u_{14}\}$, 
$\Et=\{a_1,a_2,a_3,a_4\}$, 
$\Ew=\{a_5,a_6,\ldots,a_9\}$,
$\Ez=\{a_{10}\}$,
$\Eew=\{a_{11},a_{12},\ldots,a_{18}\}$ and 
$\EC^\lnk=\{a_1,a_2\}$.

 A {\em subdivision} $S$ of $\GC$  
is a graph constructed from a seed graph $\GC$ 
according to the following rules:
\begin{enumerate}[leftmargin=*]
\item[-]
Each edge $e=uv\in \Et$ is replaced
with a $u,v$-path $P_e$ of length at least 2;

\item[-] 
Each edge $e=uv\in \Ew$ is replaced
with a $u,v$-path $P_e$ of length at least 1
(equivalently $e$ is directly used or replaced with
a $u,v$-path $P_e$ of length at least 2);

\item[-] 
Each edge $e\in \Ez$ is either used or discarded;   and

\item[-]
Each edge $e\in \Eew$ is always used directly.
\end{enumerate}

The set of link-edges in the monomer representation  $\C$ of 
an inferred polymer 
consists of edges in $\EC^\lnk\cap( \Eew\cup \Ew)$
or edges  in   paths $P_e$ for all edges $e=uv\in \EC^\lnk\cap (\Ew\cup \Et)$
in a  subdivision  $S$ of $\GC$. 
 
A target chemical graph $\C=(H,\alpha,\beta)$ will contain  $S$  as a subgraph
of the interior $H^\inte$ of $\C$.


\subsection*{Interior-specification}

A graph $H^*$ that serves as the interior $H^\inte$ of
a target chemical graph $\C$ will be constructed as follows.
First construct a subdivision  $S$ of a seed graph $\GC$ 
by replacing each edge $e=u u'\in \Et\cup\Ew$
with a pure $u,u'$-path $P_e$.
Next construct a supergraph $H^*$ of $S$ by 
attaching a leaf path $Q_v$ at each vertex $v\in \VC$ or
at an internal vertex $v\in V(P_e)\setminus\{u,u'\}$ 
of each pure $u,u'$-path $P_e$ for some edge $e=uu'\in \Et\cup\Ew$,
where possibly $Q_v=(v), E(Q_v)=\emptyset$ 
(i.e., we do not attach any new edges to $v$).
We introduce the following rules for specifying
 the size of $H^*$, the length $|E(P_e)|$  of
a pure path  $P_e$,  the length $|E(Q_v)|$ of
a   leaf path $Q_v$, the number of  leaf paths $Q_v$
and a bond-multiplicity of each interior-edge,
where we call the set of prescribed constants  
 an  {\em interior-specification}   $\sint$: 
\begin{enumerate}[leftmargin=*]
 \item[-]
  Lower and upper bounds $\nint_\LB, \nint_\UB\in \Z_+$ 
  on   the number of interior-vertices 
of a target chemical graph~$\C$. 

 \item[-]
  Lower and upper bounds $\nlnk_\LB, \nlnk_\UB\in \Z_+$ 
  on   the number of link-edges 
of a target chemical graph~$\C$ (only for polymer). 
  
\item[-] 
For each edge $e=u u'\in \Et\cup\Ew$, 
\begin{description} 
\item[]
 a lower bound $\ell_{\LB}(e)$ and 
 an upper bound $\ell_{\UB}(e)$  on the length $|E(P_e)|$ of
 a pure $u,u'$-path $P_e$. 
(For a notational convenience, set 
$\ell_\LB(e):=0$, $\ell_\UB(e):=1$, $e\in \Ez$ and
$\ell_\LB(e):=1$, $\ell_\UB(e):=1$, $e\in \Eew$.)
   
\item[]  
 a lower bound $\bl_{\LB}(e)$ and 
 an upper bound $\bl_{\UB}(e)$ on the number of leaf paths $Q_v$ attached 
 at  internal vertices $v$ of a pure $u,u'$-path $P_e$.   

\item[] 
 a lower bound $\ch_{\LB}(e)$ and 
 an upper bound $\ch_{\UB}(e)$  on the maximum 
 length  $|E(Q_v)|$ of a leaf path $Q_v$ attached  
 at an internal vertex $v\in V(P_e)\setminus\{u,u'\}$ 
 of a pure $u,u'$-path $P_e$.   
\end{description} 

\item[-]
For each vertex $v\in \VC$, 
\begin{description} 
\item[]
 a lower bound $\ch_{\LB}(v)$ and 
 an upper bound $\ch_{\UB}(v)$  on  
 the number of leaf paths $Q_v$ attached to $v$,
 where $0\leq \ch_{\LB}(v)\leq \ch_{\UB}(v)\leq 1$.
 
\item[]
 a lower bound $\ch_{\LB}(v)$ and 
 an upper bound $\ch_{\UB}(v)$  on the
 length $|E(Q_v)|$ of a leaf path $Q_v$ attached to $v$. 
\end{description}  

\item[-] 
For each edge $e=u u'\in \EC$, 
a lower bound $\bd_{m, \LB}(e)$ 
and an  upper bound $\bd_{m, \UB}(e)$  on
the number of edges with bond-multiplicity $m\in [2,3]$ in
$u,u'$-path $P_e$, where we regard $P_e$, $e  \in \Ez\cup \Eew$ 
as single edge $e$.
\end{enumerate}

We call a graph $H^*$ that satisfies an interior-specification $\sint$
a {\em $\sint$-extension of $\GC$}, 
where the bond-multiplicity of each edge has been determined.

Table~\ref{table:interior-spec}  shows an example of 
an interior-specification  $\sint$ to the seed graph  $\GC$ in 
Figure~5(a). 

\begin{table}[h!]\caption{Example~1 of an interior-specification  $\sint$. }
\begin{tabular}{ |  c | c | c | c |  } \hline 
$\nint_\LB=20$ & $\nint_\UB = 30$ & 
$\nlnk_\LB=2$ & $\nlnk_\UB = 24$ \\\hline 
\end{tabular}

 \begin{tabular}{ |  c | c c c c c c c c c |  } \hline
              & $a_1$ &  $a_2$ &   $a_3$ &   $a_4$ &   $a_5$ &   $a_6$ &   $a_7$ &   $a_8$  &   $a_9$   \\\hline
 $\ell_\LB(a_i)$ &  2 &  4 &  3 & 2 &  2 &  1  &  1 &  1 &   1\\ \hline
 $\ell_\UB(a_i)$ &  3 & 6 &  6 & 5 &  3 &  3  &  6 &  2 &   6 \\\hline
 $\bl_\LB(a_i)$ &   0 &  1 & 1 & 0 &  0 &   0 &  0 &   0 &  0 \\ \hline
 $\bl_\UB(a_i)$ &  1 &  4 &  4 & 3 &  2 &   1 &  1 &  1  &  1 \\\hline
 $\ch_\LB(a_i)$ &   0 &  2 &  1 & 0 &  0 &  0 &  0 &   0 &   0 \\ \hline
 $\ch_\UB(a_i)$ &  3 &  6 &  6 & 3 &  3 &   3 &  3 &   0 &   0 \\\hline
\end{tabular} 

\begin{tabular}{ |  c | c c c c c c   c c c c  c c c  c|  } \hline
                        & $u_1$ &  $u_2$ &   $u_3$ &   $u_4$ &   $u_5$ &   $u_6$ 
                       & $u_7$ &   $u_8$ &   $u_9$ &   $u_{10}$ &   $u_{11}$ 
                       &   $u_{12}$ &   $u_{13}$ &   $u_{14}$ \\\hline 
 $\bl_\LB(u_i)$ &  0 & 0 & 0 & 0 & 0 &  0 & 0 & 0 & 1 & 0 & 0 & 0 & 0 & 0 \\ \hline
 $\bl_\UB(u_i)$&  1 & 1 & 1 & 1 & 1 &  1 & 1 & 1 & 1 & 1 & 1 & 1 & 1 & 1 \\\hline
 $\ch_\LB(u_i)$&  0 & 0 & 0 & 0 & 0 &  0 & 0 & 0 & 1 & 0 & 0 & 0 & 0 & 0 \\\hline
 $\ch_\UB(u_i)$& 4 & 4 & 4 & 4 & 4 &  4 & 4 & 4 & 6 & 4 & 4 & 4 & 4 & 4 \\\hline
\end{tabular} 

\begin{tabular}{ |  c | c c c c c c   c c c c c c  c c c c c c |  } \hline
                               & $a_1$ &  $a_2$ &   $a_3$ &   $a_4$ &   $a_5$ &   $a_6$ 
                               & $a_7$ &  $a_8$ &   $a_9$ &   $a_{10}$ &   $a_{11}$ &   $a_{12}$ 
                               & $a_{13}$ &   $a_{14}$ &   $a_{15}$ &   $a_{16}$   &   $a_{17}$ &   $a_{18}$
                                    \\\hline
 $\bd_{2, \LB}(a_i)$ &  0 & 0 & 0 & 0 & 0 &  0 & 0 & 0 & 0 & 0 & 0 & 0 & 0 & 1  & 0 & 0 & 0 & 0\\ \hline 
 $ \bd_{2, \UB}(a_i)$& 1 & 2 & 1 & 1 & 1 &  1 & 1 & 1 & 1 & 1 & 1 &  1 & 1 & 1 & 1 & 1 &  1 & 1\\ \hline
 $\bd_{3, \LB}(a_i)$ &  0 & 0 & 0 & 0 & 0 &  0 & 0 & 0 & 0 & 0 & 0 & 0 & 0 & 0  & 0 & 0 & 0 & 0\\ \hline
 $ \bd_{3, \UB}(a_i)$& 1 & 1 & 1 & 1 & 1 &  1 & 1 & 1 & 1 & 1 & 1 &  1 & 1 & 1 & 1 &  1 & 1 & 1\\ \hline
\end{tabular} 
\label{table:interior-spec}  
\end{table}

Figure~6 
illustrates an example of 
an $\sint$-extension $H^*$ of seed graph  $\GC$ in 
Figure~5(a) 
under the interior-specification  $\sint$ in 
Table~\ref{table:interior-spec}.


\subsection*{Chemical-specification}
 
 Let $H^*$ be a graph that serves as 
 the interior $H^\inte$ of a target chemical graph $\C$,
 where the bond-multiplicity of each edge in $H^*$ has be determined.
 Finally we introduce a set of rules for constructing 
   a target chemical graph $\C$ from $H^*$ 
   by choosing  a chemical element $\ta\in \Lambda$ 
and assigning a ${\rho}$-fringe-tree $\psi$
 to each interior-vertex $v\in V^\inte$. 
We introduce the following rules for specifying
the size of $\C$, a set of chemical rooted trees  
that are allowed to use as  ${\rho}$-fringe-trees 
and lower and upper bounds on the frequency of
a chemical element, a chemical symbol, 
 an edge-configuration, and a fringe-configuration
where we call the set of prescribed constants   
 a  {\em chemical specification} $\sce$.
 Notice that the ones involving link-edges and connecting-vertices are only used for the inference of polymers.
 
\begin{enumerate}[leftmargin=*]
\item[-] 
Lower and upper bounds $n_\LB,  n^*\in \Z_+$
on the number of vertices, where $\nint_\LB \leq n_\LB\leq n^*$.
 
\item[-] 
A subset $\mathcal{F}^* \subseteq \mathcal{F}(D_\pi)$  
 of chemical rooted trees $\psi$ with $\h(\anpsi)\leq {\rho}$, where 
 we require that 
 every ${\rho}$-fringe-tree $\C[v]$ rooted at an interior-vertex $v$ 
    in  $\C$  belongs to $\mathcal{F}^*$.  
Let   
$\Lambda^\ex$ denote the set of  chemical elements assigned to non-root
vertices over all chemical rooted trees in $\mathcal{F}^*$.  
 
\item[-] 
A subset  $\Lambda^\inte\subseteq \Lambda^\inte(D_\pi)$, where 
 we require that every chemical element $\alpha(v)$ 
 assigned to an interior-vertex  $v$ in $\C$ belongs to $\Lambda^\inte$.
Let $\Lambda:= \Lambda^\inte\cup \Lambda^\ex$ and
 $\na_\ta(\C)$ (resp., $\na_\ta^\inte(\C)$ and $\na_\ta^\ex(\C)$) 
 denote the number of vertices   (resp.,   interior-vertices and  exterior-vertices)
  $v$ such that $\alpha(v)=\ta$   in  $\C$.
 
\item[-] 
A set $\Ldg^\inte\subseteq \Lambda\times [1,4]$  of chemical  symbols.

\item[-] 
Subsets $\Gamma^\lnk\subseteq \Gamma^\inte$ of $\Gamma^\inte(D_\pi)$  
of  edge-configurations  $(\mu,\mu' ,m)$ with $\mu \leq \mu'$, where 
 we require that the edge-configuration $\ec(e)$ of an interior-edge (resp., a link-edge) $e$ in $\C$ 
 belongs to $\Gamma^\inte$ (resp.,    $\Gamma^\lnk$).
We do not distinguish  $(\mu,\mu' ,m)$ and $(\mu' , \mu,m)$.  
  
\item[-] 
Define  $\Gac^\inte$  (resp.,    $\Gac^\lnk$)  to be the set of   adjacency-configurations such that  
$\Gac^\typ:=\{(\ta, \tb, m) \mid (\ta d, \tb d',m)\in \Gamma^\typ\}, \typ\in\{\inte,\lnk\}$.   
Let  $\ac_\nu^\inte(\C), \nu\in \Gac^\inte$  
(resp.,  $\ac_\nu^\lnk(\C), \nu\in \Gac^\lnk$)   
denote  the number of  interior-edges (resp.,  link-edges) $e$
 such that $\ac(e)=\nu$  in $\C$.
  
\item[-] 
 Subsets $\Lambda^*(v)\subseteq \{\ta\in \Lambda^\inte\mid \val(\ta)\geq 2\}$, 
 $v\in \VC$,  
 we require that every chemical element $\alpha(v)$ 
 assigned to   a vertex $v\in  \VC$
 in the seed graph  belongs to $\Lambda^*(v)$.  

\item[-] Lower and upper bound functions 
$\na_\LB,\na_\UB: \Lambda\to  [0,n^*]$  and 
$\na_\LB^\inte,\na_\UB^\inte: \Lambda^\inte\to  [0,n^*]$ 
on the number of   interior-vertices  $v$ such that  $\alpha(v)=\ta$  in $\C$. 

\item[-] Lower and upper bound functions  
$\ns_\LB^\inte,\ns_\UB^\inte: \Ldg^\inte\to  [0,n^*]$ 
  on the number of   interior-vertices $v$ such that $\cs(v)=\mu$  in $\C$.   

\item[-] Lower and upper bound functions  
$\ns_\LB^\cnt,\ns_\UB^\cnt: \Ldg^\inte\to  [0,2]$ 
  on the number of connecting-vertices $v$ such that $\cs(v)=\mu$  in $\C$.   
  
\item[-] Lower and upper bound functions  
$\ac_\LB^\inte,\ac_\UB^\inte: \Gac^\inte \to  \Z_+$ 
($\ac_\LB^\lnk,\ac_\UB^\lnk: \Gac^\lnk \to  \Z_+$)
 on the number of  interior-edges (resp., link-edges) $e$ such that $\ac(e)=\nu$  in $\C$. 

\item[-] Lower and upper bound functions  
$\ec_\LB^\inte,\ec_\UB^\inte: \Gamma^\inte \to  \Z_+$ 
(resp., $\ec_\LB^\lnk,\ec_\UB^\lnk: \Gamma^\lnk \to  \Z_+$)  
 on the number of  interior-edges  (resp., link-edges)  $e$ such that $\ec(e)=\gamma$  in $\C$.  
 
\item[-] Lower and upper bound functions  
$\fc_\LB,\fc_\UB: \mathcal{F}^*\to  [0,n^*]$ 
  on the number of   interior-vertices $v$ 
  such that $\C[v]^\fr$ is r-isomorphic to $\psi\in \mathcal{F}^*$  in $\C$.  
  
 \item[-] Lower and upper bound functions  
$\ac^\lf_\LB,\ac^\lf_\UB: \Gac^\lf \to  [0,n^*]$ 
  on the number of  leaf-edges $uv$ in $\acC$
  with adjacency-configuration $\nu$.  
\end{enumerate}
 
We call a chemical graph $\C$ that satisfies a chemical specification $\sce$
a {\em $(\sint,\sce)$-extension of $\GC$},
and denote by $\mathcal{G}(\GC, \sint,\sce)$ the set of
all $(\sint,\sce)$-extensions of $\GC$. 

Table~\ref{table:chemical_spec}  shows an example of 
a chemical-specification  $\sce$ to the seed graph  $\GC$
 in Figure~5(a). 

\begin{table}[h!]\caption{Example~2 of a chemical-specification  $\sce$.  
}
\begin{tabular}{ |  l |  } \hline
 $n_\LB=30$,  $n^* =50$. \\\hline
  branch-parameter:   ${\rho}=2$  \\\hline
\end{tabular}

\begin{tabular}{ |  l |  } \hline
 Each of sets $\mathcal{F}(v), v\in \VC$ and
 $\mathcal{F}_E$ is set to be \\
 the set $\mathcal{F}$  of chemical rooted trees $\psi$ with $\h(\anpsi)\leq {\rho}=2$
in Figure~5(b). \\\hline  
\end{tabular}

\begin{tabular}{ |  c | c |   } \hline
  $\Lambda=\{ \ttH,\ttC,\ttN,\ttO, \ttS_{(2)},\ttS_{(6)}, \ttP=\ttP_{(6)},\ttCl\}$ & 
  $\Ldg^\inte =\{ \ttC2 , \ttC3,  \ttC4, \ttN2, \ttN3, \ttO2,
    \ttS_{(2)}2,  \ttS_{(6)}3, \ttP4   \}$  
\\\hline
\end{tabular}

\begin{tabular}{ |  c | l |  } \hline
  $\Gac^{\inte}$ &
  $ \nu_1 \!=\!(\ttC   , \ttC  , 1) ,   \nu_2 \!=\!(\ttC   , \ttC  , 2) ,   
   \nu_3 \!=\!(\ttC   , \ttN  , 1) ,  \nu_4 \!=\!(\ttC  , \ttO  , 1), 
    \nu_5 \!=\! (\ttC, \ttS_{(2)}, 1),\nu_6 \!=\!(\ttC  , \ttS_{(6)}, 1), 
    \nu_7 \!=\! (\ttC  , \ttP  , 1) $  \\ \hline
\end{tabular}

\begin{tabular}{ |  c | l |  } \hline
  $\Gamma^{\inte}$ &
  $ \gamma_1 \!=\! (\ttC 2 , \ttC 2, 1) ,
    \gamma_{2} \!=\!(\ttC 2 , \ttC 2, 2),  
   \gamma_3 \!=\!(\ttC 2 , \ttC 3, 1) ,  
   \gamma_4 \!=\!(\ttC 2 , \ttC 3, 2) ,  
   \gamma_5 \!=\!(\ttC 2 , \ttC 4, 1) , 
   \gamma_6 \!=\!(\ttC 3 , \ttC 3, 1) , 
 $ \\
   &
  $  \gamma_7 \!=\!(\ttC 3 , \ttC 3, 2) ,
    \gamma_8 \!=\!(\ttC 3 , \ttC 4, 1), 
   \gamma_9 \!=\!(\ttC 2 , \ttN 3, 1) ,  
   \gamma_{10} \!=\!(\ttC 3 , \ttN 2, 1) ,   
    \gamma_{11} \!=\!(\ttC 4 , \ttN2, 1), 
   \gamma_{12} \!=\!(\ttC 2 , \ttO 2, 1), $ \\ 
   &
  $  
    \gamma_{13} \!=\!(\ttC 3 , \ttO 2, 1) ,    
    \gamma_{14} \!=\!(\ttC 2, \ttS_{(2)} 2, 1),  
    \gamma_{15} \!=\!(\ttC 3, \ttS_{(2)} 2, 1),  
    \gamma_{16} \!=\!(\ttC 4, \ttS_{(2)} 2, 1),  
    \gamma_{17} \!=\!(\ttC 3 , \ttS_{(6)}3, 1),   $ \\ 
   &
  $  
   \gamma_{18} \!=\!(\ttC 4, \ttS_{(6)}3, 1), 
    \gamma_{19} \!=\!(\ttC 2, \ttP4, 1), 
    \gamma_{20} \!=\!(\ttC 3, \ttP4, 1)  
     $ \\ \hline
\end{tabular}

\begin{tabular}{ |  c | l |  } \hline
  $\Gac^{\lnk}$ &
  $ \nu'_1 \!=\!(\ttC   , \ttC  , 1) ,   \nu'_2 \!=\!(\ttC   , \ttC  , 2) ,   
   \nu'_3 \!=\!(\ttC   , \ttN  , 1),  \nu'_4 \!=\!(\ttC   , \ttS_{(2)}  , 1)  $  \\ \hline
\end{tabular}

\begin{tabular}{ |  c | l |  } \hline
  $\Gamma^{\lnk}$ &
  $ \gamma'_1 \!=\! (\ttC 2 , \ttC 2, 1) ,
   \gamma'_2 \!=\!(\ttC 2 , \ttC 3, 1) ,  
   \gamma'_3 \!=\!(\ttC 2 , \ttC 4, 1) ,  
   \gamma'_4 \!=\!(\ttC 3 , \ttC 3, 1) , 
   \gamma'_5 \!=\!(\ttC 3 , \ttC 3, 2) ,
   \gamma'_6 \!=\!(\ttC 2 , \ttN 3, 1) ,   $ \\
   &
  $   
   \gamma'_7 \!=\!(\ttC 3 , \ttN 2, 1), 
    \gamma'_8 \!=\!(\ttC 2, \ttS_{(2)} 2, 1),  
    \gamma'_9 \!=\!(\ttC 3, \ttS_{(2)} 2, 1),  
    \gamma'_{10} \!=\!(\ttC 4, \ttS_{(2)} 2, 1) $  \\\hline
\end{tabular}

\begin{tabular}{ |  l|  } \hline
$\Lambda^*(u_i)=\{\ttC\}, i\in\{1,2,3,4,5,6,9\}$, 
$\Lambda^*(u_8)=\{{\ttO}\}$, 
$\Lambda^*(u_{12})=\{{\tt C, P}\}$, \\
$\Lambda^*(u_i)=\{\ttC,\ttO,\ttN\}$, $i\in [1,14]\setminus\{1,2,3,4,5,6,8,9,12\}$
   \\\hline
\end{tabular}
  
\begin{tabular}{ |  c | c c c c  c c c c |  } \hline
                         & ${\tt H}$  & ${\tt C}$ &   ${\tt N}$ &     ${\tt O}$ 
                         & $\ttS_{(2)}$ & $\ttS_{(6)}$ & $\ttP$ & $\ttCl$ \\\hline
 $\na_\LB(\ta)$ & 40 &  25 & 1 &  1 & 0 & 0 & 0  & 0 \\ \hline 
 $\na_\UB(\ta)$ & 80 & 50 & 8 &  8  & 4 & 4 & 4 & 4 \\\hline
\end{tabular} 
\begin{tabular}{ |  c | c c c  c c c   |  } \hline
   & $\ttC$ &   $\ttN$ &     $\ttO$  & $\ttS_{(2)}$ & $\ttS_{(6)}$ & $\ttP$  \\\hline
 $\na_\LB^{\inte}(\ta)$ &  10 &  1 & 0  & 0 & 0 & 0      \\ \hline
 $\na_\UB^{\inte}(\ta) $&  25 & 4 & 5 & 2 & 2 & 2  \\\hline
\end{tabular} 

\begin{tabular}{ |  c | c c c c c c  c c c   |  } \hline
    & $\ttC2$ &  $\ttC3$ &   $\ttC4$ & $\ttN2$ &   $\ttN3$ &   $\ttO2$
   & $\ttS_{(2)}2$ & $\ttS_{(6)}3$ & $\ttP4$  \\\hline
 $\ns_\LB^{\inte}(\mu)$ &  3 &   5 & 0  & 0 &  0 &  0 & 0 &  0 &  0    \\ \hline
 $\ns_\UB^{\inte}(\mu) $& 12 & 15 & 5 & 5 &  3 &  5  & 1 & 1 &  1   \\\hline
\end{tabular} 

\begin{tabular}{ |  c | c c c c c c  c c c   |  } \hline
    & $\ttC2$ &  $\ttC3$ &   $\ttC4$ & $\ttN2$ &   $\ttN3$ &   $\ttO2$
   & $\ttS_{(2)}2$ & $\ttS_{(6)}3$ & $\ttP4$  \\\hline
 $\ns_\LB^{\cnt}(\mu)$ &  0 &   0 & 0  & 0 &  0 &  0 & 0 &  0 &  0    \\ \hline
 $\ns_\UB^{\cnt}(\mu) $& 2 & 2 & 2 & 2 & 2 &  2  & 1 & 1 &  0   \\\hline
\end{tabular}   

\begin{tabular}{ |  c | c c c c c c c |  } \hline
         & $\nu_1 $ &   $\nu_2 $ & $\nu_3 $   & $\nu_4 $
         &   $\nu_5 $ & $\nu_6 $   & $\nu_7 $ \\\hline
 $\ac_\LB^{\inte}(\nu)$  &  0  &  0  & 0  &  0  & 0  &  0 & 0     \\ \hline
 $\ac_\UB^{\inte}(\nu)$ & 30 & 10 & 10 &  10 & 2 &  3 &  3 \\\hline
\end{tabular} 

\begin{tabular}{ |  c | c c c c c  c c  |  } \hline
    & $\gamma_1 $ &   $\gamma_2 $ & $\gamma_3 $   & $\gamma_4 $  & $\gamma_5 $
    & $\gamma_i, i\in[6,13]$ &   $\gamma_i, i\in[14,20]$    \\\hline
 $\ec_\LB^{\inte}(\gamma)$ &
    0 &  0 & 0 &  0  & 0 &  0 &  0    \\ \hline
 $\ec_\UB^{\inte}(\gamma) $ &
   4 & 15 & 5 &  5  & 10 & 5  &   2  \\\hline
\end{tabular} 

\begin{tabular}{ |  c | c c c  c  |  } \hline
         & $\nu'_1 $ &   $\nu'_2 $ & $\nu'_3 $  & $\nu'_4 $  \\\hline
 $\ac_\LB^{\lnk}(\nu')$  &  0  &  0  & 0  & 0      \\ \hline
 $\ac_\UB^{\lnk}(\nu')$ &  10 &  5 &  5  &  5   \\\hline
\end{tabular} 
\begin{tabular}{ |  c | c    |  } \hline
    & $\gamma'_i, i\in[1,10]$   \\\hline
 $\ec_\LB^{\lnk}(\gamma')$ &  0    \\ \hline
 $\ec_\UB^{\lnk}(\gamma') $& 4   \\\hline
\end{tabular}

\begin{tabular}{ |  c | c   c   |  } \hline 
& $\psi\in\{\psi_i\mid i=1,6,11\}$ 
& $\psi\in \mathcal{F}^*\setminus \{\psi_i\mid i=1,6,11\}$ \\\hline
 $\fc_\LB(\psi)$  &  1 &    0   \\ \hline 
 $\fc_\UB(\psi)$ &  10 &  3\\\hline
\end{tabular} 

\begin{tabular}{ |  c | c   c   |  } \hline 
& $\nu\in\{(\ttC,\ttC,1),(\ttC,\ttC,2)  \}$ 
& $\nu\in \Gac^\lf \setminus \{(\ttC,\ttC,1),(\ttC,\ttC,2)  \}$   \\\hline
 $\ac^\lf_\LB(\nu)$  &  0 &    0   \\ \hline 
 $\ac^\lf_\UB(\nu)$ &  10 &  8 \\\hline
\end{tabular} 

\label{table:chemical_spec}
\end{table}

Figure~3 
 illustrates an example of 
a   $(\sint,\sce)$-extension of $\GC$   obtained 
from the  $\sint$-extension $H^*$  
 in Figure~6 
under the chemical-specification $\sce$ in Table~\ref{table:chemical_spec}.  
  

\section{Test Instances for Phase~2}\label{sec:test_instances} 

We prepared the following instances $I_{\mathrm{a}}$ and $I_{\mathrm{b}}$
 for conducting experiments
of  Stages~4  and 5 in Phase~2
for the two data sets $D_\chi \in \{$\chiaoki, \chinistane $\}$,
and instances $I_{\mathrm{c1}}$ and $I_{\mathrm{c2}}$
of Stage~4 for the data set \chijsol.
 

\begin{figure}[t!]
\begin{center} 
 \includegraphics[width=.89\columnwidth]{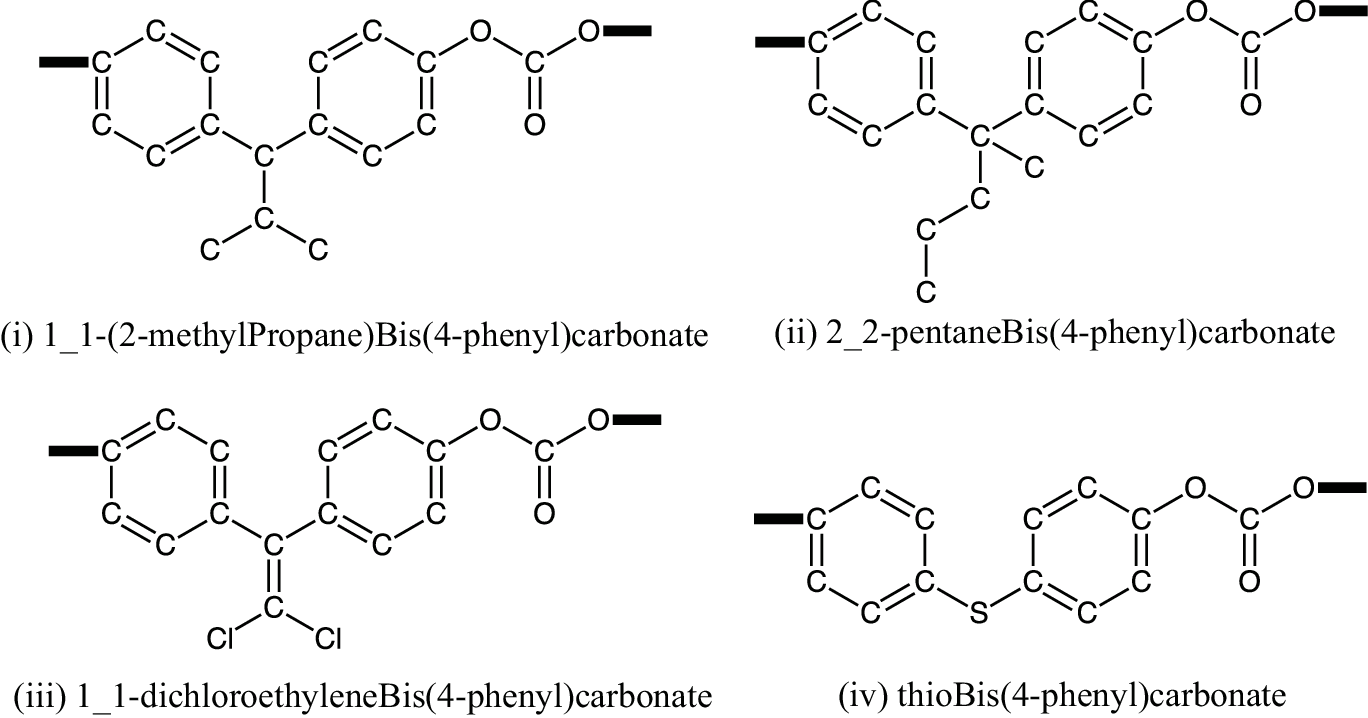}
\end{center}
\caption{Illustrations of  four polymers: 
(i)  1$\underline{~}$1-(2-methylPropane)Bis(4-phenyl)carbonate;
(ii) 2$\underline{~}$2-pentaneBis(4-phenyl)carbonate;
(iii)  1$\underline{~}$1-dichloroethyleneBis(4-phenyl)carbonate;
(iv) thioBis(4-phenyl)carbonate.
Hydrogens are omitted and connecting-edges are depicted with thick lines. 
The figure is adapted from~\cite{Ido:2024aa}.}
\label{fig:four_polymers}  
\end{figure} 
 
\begin{itemize} 
  \item[(a)]  $I_{\mathrm{a}} =(\GC,\sint,\sce)$: The instance
  used in Appendix~\ref{sec:specification} to explain the topological specification.
 
 \end{itemize}
 
\begin{itemize} 
  \item[(b)]  $I_{\mathrm{b}} =(\GC,\sint,\sce)$: An instance that
  represents  a set of polymers that includes the four examples of polymers in 
Figure~\ref{fig:four_polymers}.
We set a seed graph  $\GC=(\VC,\EC=\Eew)$ to be the graph  
 with two cycles $C_1$ and $C_2$ in Figure~8(i), 
 where we set  
$\Et=\EC^\lnk=\{a_1,a_2\}$ and 
$\Eew=\{a_{3},a_{12},\ldots,a_{14}\}$. \\
Set  $\Lambda:= \{ \ttH, \ttC, \ttN, \ttO, \ttS_{(2)}, \ttCl \}$ for each data set $D_\chi \in  \{$\chiaoki, \chinistane $\}$,
and  set $\Ldg^\inte$ to be
the set of all possible chemical symbols in $\Lambda\times[1,4]$.\\
Set 
$\Gamma^\inte$ (resp.,  $\Gamma^\lnk$)
to be the set of the edge-configurations of the interior-edges
(resp.,  the link-edges)
used in the four examples of polymers in Figure~7. 
Set 
$\Gamma^\inte_\ac$ (resp.,  $\Gamma^\lnk_\ac$) to be
 the set of the adjacency-configurations of the edge-configurations in 
$\Gamma^\inte$ (resp.,  $\Gamma^\lnk$). \\
We specify $n_\LB:=25$ 
and 
set 
$\nint_\LB:=14$, $\nint_\UB:=n^*:=n_\LB+10$,  
$\nlnk_\LB:=2$,  $\nlnk_\UB: =2+\max\{ n_\LB-15, 0\}$.  \\
For each link-edge $a_i\in\Et=\EC^\lnk=\{a_1,a_2\}$, 
set 
 $\ell_\LB(a_i):=2+\max\{\lfloor (n_\LB-15)/4\rfloor,0\}$,  
 $\ell_\UB(a_i):=\ell_\LB(a_i)+5$,
 $\bl_\LB(a_i):=0, \bl_\UB(a_i):=3$, 
 $\ch_\LB(a_i):=0, \ch_\UB(a_i):=5$,  
 $\bd_{2,\LB}(a_i):=0$ and $\bd_{2,\UB}(a_i):= \lfloor \ell_\LB(a_i)/3 \rfloor$.\\
To form two benzene rings from the two cycles $C_1$ and $C_2$, set 
   $\Lambda^*(u):=\{{\tt C}\}$, 
 $\bl_\LB(u):=\bl_\UB(u):=\ch_\LB(u):=\ch_\UB(u):=0$, $u\in \VC$,
 $\bd_{2,\LB}(a_i):=\bd_{2,\UB}(a_i):=0,  i\in\{3,5,7,9,11,13\}$,
 $\bd_{2,\LB}(a_i):=\bd_{2,\UB}(a_i):=1,  i\in\{4,6,8,10,12,14\}$.\\
Not to include any triple-bond, set 
 $\bd_{3,\LB}(a):=\bd_{3,\UB}(a):=0, a\in \EC$.
 \\
Set lower bounds
 $\na_\LB$,  $\na^\inte_\LB$,  $\ns^\inte_\LB$,  $\ns^\cnt_\LB$, 
$\ac^\inte_\LB$, $\ac_\LB^\lnk$, $\ec_\LB^\inte$, $\ec_\LB^\lnk$ and  $\ac^\lf_\LB$  to be 0. \\
Set  upper bounds   
 $\na_\UB(\ta):=n^*, \ta\in\{\ttH,\ttC\}$,   
 $\na_\UB(\ta):=5+\max\{ n_\LB-15, 0\}, \ta\in\{\ttO,\ttN\}$,
 $\na_\UB(\ta):=2+\max\{\lfloor (n_\LB-15)/4\rfloor,0\}, 
 \ta\in\Lambda\setminus \{\ttH,\ttC,\ttO,\ttN\}$,  
  $\ns^\cnt_\UB(\mu):=2, \mu\in\Ldg^\inte$, 
 and   $\na^\inte_\UB$,  $\ns^\inte_\UB$, 
$\ac^\inte_\UB$, $\ac_\LB^\lnk$, $\ec_\UB^\inte$,  $\ec_\UB^\lnk$ 
and  $\ac^\lf_\UB$ to be  $n^*$. \\
Set $\mathcal{F}$ to be the set of the 17 chemical rooted trees $\psi_i,i\in[1,17]$
 in  Figure~8(ii). 
Set $\mathcal{F}_E :=\mathcal{F}(v) := \mathcal{F}$, $v\in \VC$ and 
$\fc_\LB(\psi):=0, \psi\in \mathcal{F}$,
$\fc_\UB(\psi_i):=12+\max\{ n_\LB-15, 0\}, i\in[1,4]$, 
$\fc_\UB(\psi_i):=8+\max\{\lfloor (n_\LB-15)/2\rfloor,0\}, i\in[5,12]$ and
$\fc_\UB(\psi_i):=5+\max\{\lfloor (n_\LB-15)/4\rfloor,0\}, i\in[13,17], \psi_i\in \mathcal{F}$. 

\item[(c1)] $I_{\mathrm{c1}}=(\GC,\sint,\sce)$: An instance that represents a relatively simple polymer structure.
The seed graph is illustrated in Figure~\ref{fig:figure_c1c2}(a),
where we set  
$\Ew=\{a_1 \}$, $\Eew=\{ a_2, a_3\}$, and $\EC^\lnk=\{ a_1, a_2, a_3 \}$. \\
Set  $\Lambda:= \{ \ttH, \ttC, \ttN, \ttO, \ttCl \}$ 
and  set $\Ldg^\inte$ to be
the set of all possible chemical symbols in $\Lambda\times[1,4]$.\\
Set 
$\Gamma^\inte$ (resp.,  $\Gamma^\lnk$)
to be the set of all the edge-configurations of the interior-edges
(resp.,  the link-edges)
appeared in the data set \chijsol.
Set 
$\Gamma^\inte_\ac$ (resp.,  $\Gamma^\lnk_\ac$) to be
 the set of all the adjacency-configurations of the edge-configurations in 
$\Gamma^\inte$ (resp.,  $\Gamma^\lnk$). \\
We 
set $n_\LB:=\nint_\LB:=\nlnk_\LB:=3$,
$n^*:=\nint_\UB:=\nlnk_\UB:=10$.  \\
For each link-edge $a_i\in\Et=\EC^\lnk=\{a_1,a_2,a_3\}$, 
set 
 $\ell_\LB(a_i):=1$,  
 $\ell_\UB(a_1):=5$, $\ell_\UB(a_2):=\ell_\UB(a_3):=1$
 $\bl_\LB(a_i):=0, \bl_\UB(a_i):=0$, 
 $\ch_\LB(a_i):=0, \ch_\UB(a_1):=1,  \ch_\UB(a_2):=0,  \ch_\UB(a_3):=2$,  
 $\bd_{2,\LB}(a_i):=0$ and $\bd_{2,\UB}(a_1):= 1, \bd_{2,\UB}(a_2):= 3, \bd_{2,\UB}(a_3):= 0$.\\
Not to include any triple-bond, set 
 $\bd_{3,\LB}(a):=\bd_{3,\UB}(a):=0, a\in \EC$.
 \\
Set lower bounds
 $\na_\LB$,  $\na^\inte_\LB$,  $\ns^\inte_\LB$,  $\ns^\cnt_\LB$, 
$\ac^\inte_\LB$, $\ac_\LB^\lnk$, $\ec_\LB^\inte$, $\ec_\LB^\lnk$ and  $\ac^\lf_\LB$  to be 0. \\
Set  upper bounds   
 $\na_\UB(\ta):=n^*, \ta\in\{\ttH,\ttC,\ttN,\ttO\}$,   
 $\na_\UB(\ta):=5, \ta=\ttCl$,
  $\ns^\cnt_\UB(\mu):=2, \mu\in\Ldg^\inte$, 
  $\ac^\lf_\UB(\nu):=5, \nu\in\Gac^\lf$,
 and   $\na^\inte_\UB$,  $\ns^\inte_\UB$, 
$\ac^\inte_\UB$, $\ac_\LB^\lnk$, $\ec_\UB^\inte$,  and  $\ec_\UB^\lnk$ 
 to be  $n^*$. \\
Set $\mathcal{F}$ to be the set of  all the chemical rooted trees that appeared in the data set \chijsol.
Set $\mathcal{F}_E :=\mathcal{F}(v) := \mathcal{F}$, $v\in \VC$ and 
$\fc_\LB(\psi):=0, \fc_\UB(\psi):=n^*, \psi\in \mathcal{F}$.

\item[(c2)] $I_{\mathrm{c2}}=(\GC,\sint,\sce)$: An instance that represents a polymer structure with a benzene ring included.
The seed graph is illustrated in Figure~\ref{fig:figure_c1c2}(b),
where we set  
$\Ew=\{a_1, a_2 \}$, $\Eew=\{ a_3, a_4, ..., a_{10}\}$, and $\EC^\lnk=\{ a_1, a_2, a_9, a_{10} \}$. \\
Set  $\Lambda:= \{ \ttH, \ttC, \ttN, \ttO, \ttCl \}$ 
and  set $\Ldg^\inte$ to be
the set of all possible chemical symbols in $\Lambda\times[1,4]$.\\
Set 
$\Gamma^\inte$ (resp.,  $\Gamma^\lnk$)
to be the set of all the edge-configurations of the interior-edges
(resp.,  the link-edges)
appeared in the data set \chijsol.
Set 
$\Gamma^\inte_\ac$ (resp.,  $\Gamma^\lnk_\ac$) to be
 the set of all the adjacency-configurations of the edge-configurations in 
$\Gamma^\inte$ (resp.,  $\Gamma^\lnk$). \\
We 
set $n_\LB:=\nint_\LB:=10$, $\nlnk_\LB:=10$,
$\nint_\UB:=\nlnk_\UB:=15$, and $n^*:=25$.  \\
For each link-edge $a_i\in\Et=\EC^\lnk=\{a_1,a_2,a_9,a_{10}\}$, 
set 
 $\ell_\LB(a_i):=1$,  
 $\ell_\UB(a_1):=\ell_\UB(a_2):=5$, $\ell_\UB(a_9):=\ell_\UB(a_{10}):=1$
 $\bl_\LB(a_i):=0, \bl_\UB(a_i):=1$, 
 $\ch_\LB(a_i):=0, \ch_\UB(a_1):=\ch_\UB(a_2):=3$,  
 $\bd_{2,\LB}(a_i):=0$ and $\bd_{2,\UB}(a_1):= 1, \bd_{2,\UB}(a_2):=\bd_{2,\UB}(a_9):= 0, \bd_{2,\UB}(a_{10}):= 3$.\\
 To contain a benzene ring, set
   $\Lambda^*(u):=\{{\tt C}\}$, 
 $\bl_\LB(u):=\bl_\UB(u):=\ch_\LB(u):=\ch_\UB(u):=0$, $u\in \{u_2, u_3, ..., u_7  \}$,
 $\bd_{2,\LB}(a_i):=\bd_{2,\UB}(a_i):=0,  i\in\{4,6,8\}$,
 $\bd_{2,\LB}(a_i):=\bd_{2,\UB}(a_i):=1,  i\in\{3,5,7\}$.\\
Not to include any triple-bond, set 
 $\bd_{3,\LB}(a):=\bd_{3,\UB}(a):=0, a\in \EC$.
 \\
Set lower bounds
 $\na_\LB$,  $\na^\inte_\LB$,  $\ns^\inte_\LB$,  $\ns^\cnt_\LB$, 
$\ac^\inte_\LB$, $\ac_\LB^\lnk$, $\ec_\LB^\inte$, $\ec_\LB^\lnk$ and  $\ac^\lf_\LB$  to be 0. \\
Set  upper bounds   
 $\na_\UB(\ta):=n^*, \ta=\ttH$,   
  $\na_\UB(\ta):=20, \ta=\ttC$,   
   $\na_\UB(\ta):=10, \ta\in\{\ttN,\ttO\}$,   
 $\na_\UB(\ta):=5, \ta=\ttCl$,
  $\ns^\cnt_\UB(\mu):=2, \mu\in\Ldg^\inte$, 
  $\ac^\lf_\UB(\nu):=5, \nu\in\Gac^\lf$,
   $\na^\inte_\UB(\ta):=20, \ta=\ttC$,   
   $\na^\inte_\UB(\ta):=10, \ta\in\{\ttN,\ttO\}$,   
 $\na^\inte_\UB(\ta):=5, \ta=\ttCl$,
 $\ns^\inte_\UB(\mu):=20, \mu\in\{\ttC2,\ttC3,\ttC4\}$,
 $\ns^\inte_\UB(\mu):=10, \mu\in\Ldg^\inte\setminus\{\ttC2,\ttC3,\ttC4\}$,
 and   
$\ac^\inte_\UB$, $\ac_\LB^\lnk$, $\ec_\UB^\inte$,  and  $\ec_\UB^\lnk$ 
 to be  20. \\
Set $\mathcal{F}$ to be the set of  all the chemical rooted trees that appeared in the data set \chijsol.
Set $\mathcal{F}_E :=\mathcal{F}(v) := \mathcal{F}$, $v\in \VC$ and 
$\fc_\LB(\psi):=0, \fc_\UB(\psi):=10, \psi\in \mathcal{F}$.

 \end{itemize}

\end{document}